\documentclass[letterpaper,twocolumn,10pt]{article}
\usepackage{usenix}

\usepackage{tikz}
\usepackage{amsmath}
\usepackage{amsmath,booktabs}
\usepackage{amssymb}
\usepackage{algorithm}
\usepackage{algorithmicx}
\usepackage{algpseudocode}
\usepackage{enumitem}
\usepackage{booktabs}
\usepackage{cleveref}
\usepackage[table]{xcolor} 
\usepackage{multirow} 
\usepackage{array}  
\usepackage{makecell}
\usepackage{graphicx}
\usepackage{subcaption}
\floatname{algorithm}{Algorithm}

\usepackage{cleveref}
\usepackage{makecell}
\usepackage{xcolor}

\usepackage{siunitx}
\usepackage{xurl}
\crefname{section}{§}{§§}
\Crefname{section}{§}{§§}
\usepackage[available]{usenixbadges}
\begin{document}

\date{}

\title{\Large \bf Turn Your Face Into An Attack Surface: \\ Screen Attack Using Facial Reflections in Video Conferencing}

\author{
{\rm Yong Huang\textsuperscript{1}, Yanzhao Lu\textsuperscript{1}, Mingyang Chen\textsuperscript{1}, En Zhang\textsuperscript{1}, Jiazi Li\textsuperscript{1}, and Wanqing Tu\textsuperscript{2}\thanks{Corresponding author.}}\\
\textit{\textsuperscript{1}Zhengzhou University, \textsuperscript{2}Durham University}\\
yonghuang@zzu.edu.cn, yanzhaolu@gs.zzu.edu.cn, chenmingyangzzu@gmail.com,\\
\{zzu202278030528, ljz202278040309\}@stu.zzu.edu.cn, wanqing.tu@durham.ac.uk
}

\maketitle

\begin{abstract}

In video conferencing, human faces serve as the primary visual focal points, playing multifaceted roles that enhance visual communication and emotional connection. 
However, we argue that a human face is also a side channel, which can unwittingly leak on-screen information through online video feeds. 
To demonstrate this, we conduct feasibility studies, which reveal that, illuminated by both ambient light and light emitted from displays, the human face can reflect optical variations of different on-screen content. 
The paper then proposes FaceTell, a novel side-channel attack system that eavesdrops on fine-grained application activities from pervasive yet subtle facial reflections during video conferencing.
We implement FaceTell in a real-world testbed with three different brands of laptops and four mainstream video conferencing platforms. FaceTell is then evaluated with 24 human subjects across 13 unique indoor environments.
With more than 12 hours of video data, FaceTell achieves a high accuracy of 99.32\% for eavesdropping on 28 popular applications and is resilient to many practical impact factors.
Finally, potential countermeasures are proposed to mitigate this new attack.

\end{abstract}

\section{Introduction}

Since the recent COVID-19 pandemic, video conferencing has become an essential tool for remote work, education, social connections, etc.
It is reported that the count of daily active Zoom users has reached 300 million and yielded over 3.3 trillion meeting minutes in 2024~\cite{Kumar2024}.
However, with the unprecedented growth in video calls, participants often multitask, such as reading news, checking email, and using chatting applications, during conferencing to combat Zoom fatigue or boost productivity~\cite{riedl2022stress}. 
The frequent engagement in these secondary activities introduces a new vulnerability, allowing untrusted participants to launch screen attacks on victims' devices remotely. 
Under such attacks, a malicious party can infer private and confidential information on victims' screens based on \textit{compromising reflections}, i.e., unintended reflections of light from screens on nearby objects.

While compromising reflections have a long history in the literature, the impact of such threats on emerging video conferencing remains understudied.
Previous attacks show that reflections on nearby objects, such as teapots, spoons, eyeglasses, and even eyes, are exploitable using either high-end telescopic lenses~\cite{backes2008compromising,backes2009tempest} or inexpensive commodity cameras~\cite{raguram2011ispy,xu2013seeing} in close proximity.
The variation in the overall luminance (color or brightness) of the background area in the webcam’s Field of View (FoV) is also leveraged to detect on-screen images ~\cite{weinberg2011still}.
Recent work~\cite{long2023private,wasswa2022proof} breaks the distance limitation by exploiting eyeglass reflections to infer screen content via a remote webcam in video conferencing.
However, these existing attacks are faced with at least one of the following problems. 
First, the popular virtual background feature in mainstream video conferencing platforms (e.g., Zoom, Teams) can blur both the background and any reflective items.
Second, the low image resolution of webcams and the variable network bandwidth render it difficult to infer screen content from participants' eyes.
Third, specular reflections (e.g., from eyeglasses) require the observer to be positioned precisely on the propagation path of the reflected light that has the same angle as the incident light~\cite{balzer2010principles}. 
Since online meeting participants do not remain stationary all the time, as we will analyze in~\cref{subsec-impact_factor}, eyeglasses do not constantly reflect light from the screen.
Moreover, not everyone wears eyeglasses in video conferencing, making eyeglasses an unreliable source for reflection attacks.
The above reasons limit the applicability of the existing screen attacks in real-world video conferencing settings.

This paper pushes the boundary of screen attacks during video conferencing by exploiting the more pervasive yet unremarkable facial reflections of screen content.
We observe that the human face, the key part of video conferencing images, is illuminated by not only ambient light but also the screen's light, making it a promising stable source of compromising reflections.
Although the face is not a polished surface, it generally exhibits a mixture of diffuse and specular reflections~\cite{egger20203d}.
Due to different angular and distance characteristics, different facial regions, such as the right and left cheeks, the forehead, and the nose, are sensitive to light from different parts of the entire screen (analyzed in~\cref{sec: feasibility study}). 
In this way, facial reflections included in online video streams can be employed to characterize applications or software with unique user interface (UI) layouts on the screen. This creates a potential attack surface for eavesdropping on users' multitasking activities during video conferencing.

This paper proposes FaceTell, a novel eavesdropping system, to leverage subtle facial reflections in online video streams to infer the secondary activities of video conferencing participants. 
More specifically, a malicious online meeting participant may launch such attacks by sending the video streams of other users to FaceTell, which will then extract facial images to support screen content prediction. 
In designing this system, we tackle the following challenges.

1) \textit{How to obtain high-quality face images from dynamic online video streams?}
Due to limitations in camera capability and variable network bandwidth, the video resolution of online meetings is variable, which typically ranges from 360p to 720p, significantly lower than that of high-end telescopic lens photos.
Moreover, human faces occupy only a small portion of the camera's field of view (FoV) (analyzed in~\cref{subsec-segmentaiton-and-reconstruction}), making it challenging to characterize the subtle facial reflection of screen content.
To handle this challenge, FaceTell extracts video frames from video conferencing streams and crops raw face images (i.e., eliminates background items) by employing Mask-R-CNN~\cite{he2017mask} and Haar feature classifier~\cite{viola2001rapid}. Then, using the CAMixerSR network~\cite{wang2024camixersr}, FaceTell implements super-resolution reconstructions of the cropped face images. Finally, reconstructed images will be resized and normalized into higher resolutions with fixed sizes.

2) \textit{How to accurately infer screen content from subtle facial reflections?}
As mentioned, it is still non-trivial to extract information from facial reflections as they are a mixture of specular and diffuse reflections~\cite{egger20203d}. 
To address this difficulty, we formulate screen content inference as a classification task. 
This classification task enhances the high-level features related to facial reflections by a residual convolutional block (ResBlock)~\cite{he2016deep} and a convolutional block attention module (CBAM)~\cite{woo2018cbam}, and devises an effective two-tier classification model for inferring on-screen content categories and applications. 
The classification results are further improved by a new heuristic algorithm that conducts label correction based on the temporal continuity of human-computer interactions.

\textbf{Experimental Results.} 
We implement FaceTell in a real-world testbed with three
different brands of laptops (i.e., ASUS, Lenovo, HP) and four mainstream video conferencing platforms, including Zoom, Teams, Skype, and WeChat. 
More than 12 hours of video data from 24 online meeting participants are collected. These 24 participants interact with 28 common computer applications in 13 different environments. 
The evaluation results demonstrate that FaceTell achieves an accuracy of 99.32\% in application prediction.
The accuracy is achieved with a runtime of approximately 124~ms per video frame.
Moreover, the experimental results show that FaceTell is robust to impact factors, including subject gender, facial occlusions, distances, angles, and ambient light.

\textbf{Contributions.} 
Our contributions are summarized below:
\begin{itemize}
    \item \textbf{A New Side-Channel Attack.} This paper is among the first to develop theoretical modeling and real-world experimental analyses to prove the feasibility of a new side-channel attack. This attack exploits subtle facial reflections in online video streams to eavesdrop on the secondary activities of video conference participants.  
    
    \item \textbf{The Novel FaceTell System.} This paper proposes a novel eavesdropping system, FaceTell, that provides an efficient tool to launch this new side-channel attack. The FaceTell system constitutes a set of novel schemes, including a two-tier classification model and a heuristic label correction algorithm, to accurately reveal victims' private information regarding their secondary activities. 
    
    \item \textbf{Comprehensive Evaluation.} FaceTell is evaluated with 28 popular computer applications, four laptops, four mainstream video conferencing platforms, and 24 victims in 13 different environments. The experimental results demonstrate that FaceTell is highly effective and robust to various potential confounding factors.
\end{itemize}


\begin{figure}[t]
    \centering
    \includegraphics[width=\linewidth]{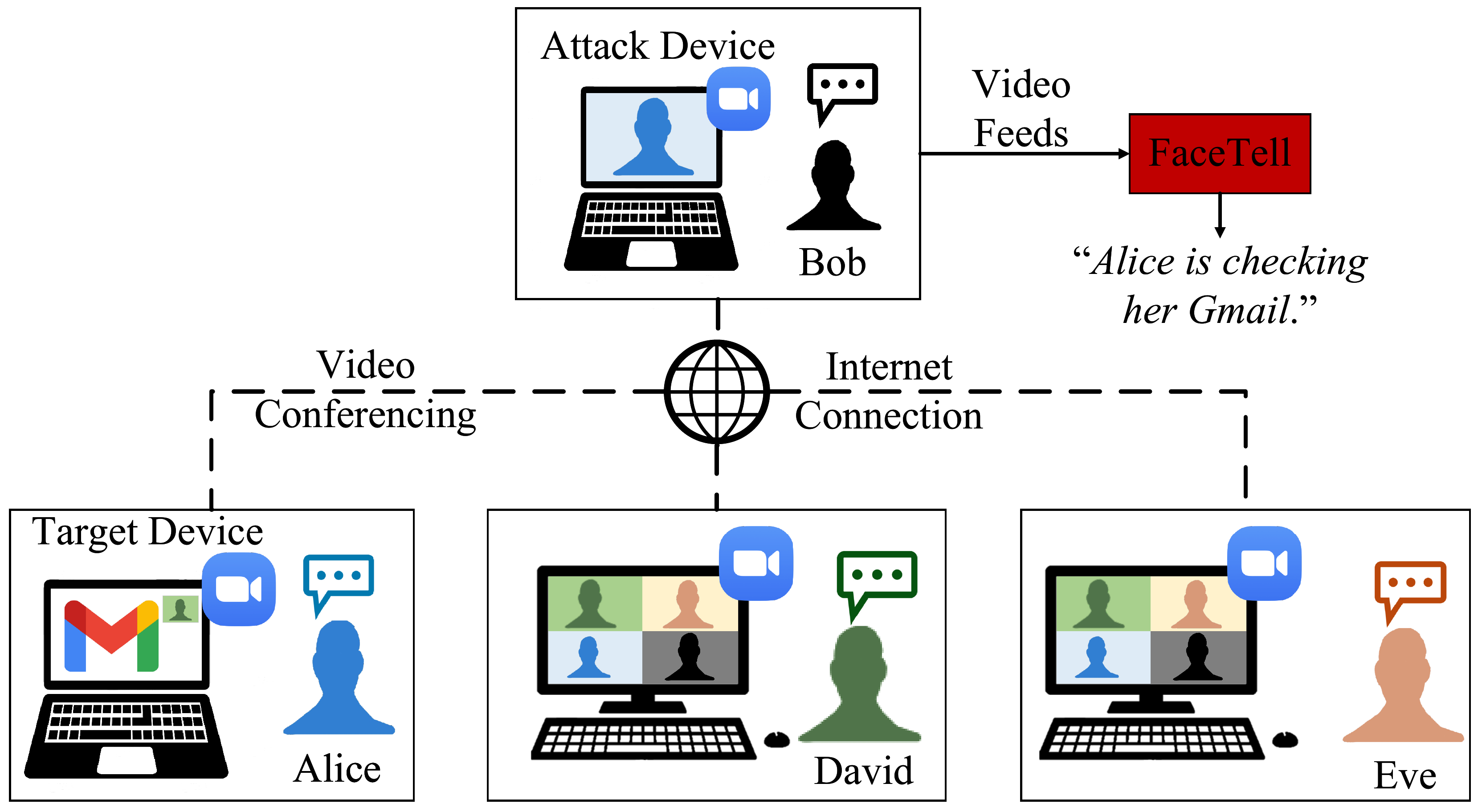}
    \caption{An example of screen-based attacks.}
    \label{fig-Threat model}
\end{figure}

\section{Threat Model}\label{sec: threat model}

A typical screen-based attack in video conferencing is illustrated in Figure~\ref{fig-Threat model}. 
Suppose that all conference participants use the same video conferencing platform (e.g., Zoom, Teams). 
As the mutual trust among video conferencing participants is not always guaranteed~\cite{cecconello2019skype}, we assume Bob is a screen-based attacker who wants to eavesdrop on the screen activities of Alice, i.e., the victim.
Hereinafter, Alice's computer is referred to as the target device, and Bob's computer is called the attack device.
Bob's goal is to identify on-screen activities by analyzing the light patterns reflected on Alice's face in available online video streams from the target device. 
Bob can achieve this goal by intercepting the video feeds from the video conferencing platform or recording these video feeds using screen recording software on the attack device.

\section{Feasibility Study}\label{sec: feasibility study}
\vspace{0.1ex}

\subsection{Theoretical Modeling}\label{subsec-theoretical_model}

\begin{figure}[t]
	\centering
	\includegraphics[width=0.9\linewidth]{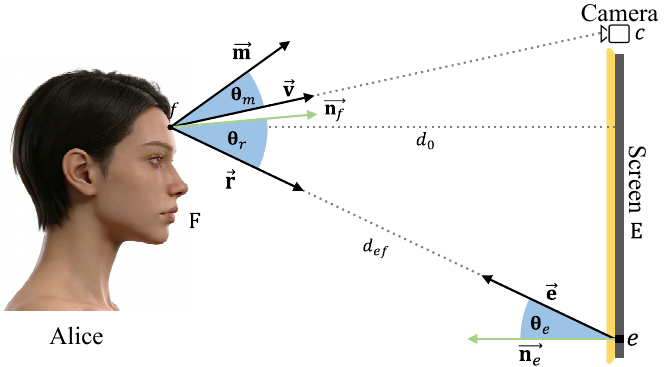}
	\caption{An example to demonstrate the theoretical model of facial reflections in video conferencing.
	}
	\label{fig:physical model}
\end{figure}

We first introduce the notations that we employ for developing the theoretical model. Let us denote the location of a digital screen as $\mathbf{E}$ as depicted in Figure~\ref{fig:physical model}.
An individual light-emitting point on this screen, such as a liquid crystal (LC) unit that acts as a light source, is located at a point $e \in \mathbf{E} $. 
The orientation of the screen at the location $e$ is defined by its normal vector $ \vec{\mathbf{n}_e} $.
The surface of Alice's face is represented by the coordinate set $\mathbf{F} $. For a single point on the face, we use $f$ to represent its coordinate, where $f \in \mathbf{F}$. Similarly, the normal vector for any point $f$ on the face is denoted as $ \vec{\mathbf{n}_f}$.
The camera is located at the location $c$.
Moreover, we define several key direction vectors to describe the path of light propagation. 
The emitting direction, representing the light ray traveling from the screen unit at $e$ to the face point at $f$, is denoted as $ \vec{ \mathbf{e} } = \frac{\vec{ef}}{\vert \vec{ef}\vert}$. 
The viewing direction at the camera location $c$ from the face point at $f$ is $ \vec{ \mathbf{v} } = \frac{\vec{fc}}{\vert \vec{fc}\vert}$. The receiving direction, from the face point at $f$ toward the screen unit at $e$, is defined as $ \vec{ \mathbf{r} } = \frac{\vec{fe}}{\vert \vec{fe}\vert}$.
In addition, $ \vec{ \mathbf{m} }$ is the mirror direction in which a light ray from the screen unit at $e$ is perfectly reflected off  $f$. 

Based on the above notations, we first deduce the light intensity from the screen unit to the face point.
The light intensity emitted by the screen unit along $ \vec{\mathbf{n}_e} $ is denoted as $I_e$, which is determined by the screen content.
Due to the spatial directivity of each LC unit, the light intensity emitted to each angle is different~\cite{svilainis2008led, svilainis2010numerical}.
Concretely, the light intensity from the screen unit to the face point can be calculated as~\cite{radiometry2001}
\begin{align}\label{eq:received intensity}
    I_f =  \frac{I_e \cdot W(\theta_e)}{(d_{ef})^2}, 
\end{align}
where $\theta_e$ is the angle between the emitting direction $ \vec{ \mathbf{e} }$ and the normal vector $\vec{\mathbf{n}_e}$, and $d_{ef}$ is the distance between $e$ and $f$.
In addition, $W(\cdot)$ represents the angular distribution of light intensity from the screen unit at $e$.

Then, we proceed to compute the optical reflection intensity of the face point at $f$ from the entire screen.
To do this, we must sum up all the LC units' light emanations.
According to the Phong reflection model~\cite{phong1998illumination}, the reflected light intensity pointing to the viewing direction $ \vec{ \mathbf{v} }$ can be expressed as
\begin{align}\label{eq:reflected inrensity}
    I_{\vec{\mathbf{V}}} =  \sum_{e \in \mathbf{E}} I_f \cdot k_d \cdot \cos \theta_r + \sum_{e \in \mathbf{E}} I_f \cdot k_s \cdot \cos^{n_s} \theta_m  + k_a \cdot I_a,
\end{align}
where $k_d$, $k_s$, and $k_a$ are the diffuse, specular, and ambient reflection coefficients, respectively.
As illustrated in Figure~\ref{fig:physical model}, $\theta_r$ is the angle between the receiving direction $ \vec{ \mathbf{r} }$ and the normal vector $ \vec{\mathbf{n}_f}$, and $\theta_m$ is the angle between the mirror direction $ \vec{ \mathbf{m} }$ and the viewing direction $ \vec{ \mathbf{v} }$.
The parameter $n_s$ is the shininess exponent, which controls the size and intensity of the specular highlight; a higher $n_s$ value results in a smaller highlight, simulating a glossier surface. The term $\cos^{n_s} \theta_m$ thus models how the specular reflection intensity falls off as the viewing direction diverges from the perfect mirror direction. Additionally, $I_a$ stands for the intensity of ambient light sources.
Therefore, in Eq.~\eqref{eq:reflected inrensity}, the first part $\sum_{e \in \mathbf{E}} I_f \cdot k_d \cdot \cos \theta_r$ corresponds to the diffuse reflection, and the second part $\sum_{e \in \mathbf{E}} I_f \cdot k_s \cdot \cos^{n_s} \theta_m$ is for the specular reflection.

Next, we integrate Eq.~\eqref{eq:received intensity} into Eq.~\eqref{eq:reflected inrensity}.
Let us assume that the screen is flat and each LC unit has the same normal vector.
Given any screen unit at $e$, its distance to a face point at $f$ can be expressed as $ d_{ef} = \frac{d_0}{\sin (\pi/2 - \theta_e)} = \frac{d_0}{\cos \theta_e} $, where $d_0$ is the perpendicular line between $f$ and $\mathbf{E}$.
In this way, we obtain a new form of the reflected light intensity $ I_{\vec{\mathbf{V}}}$ as
\begin{align}\label{eq:revised reflected inrensity}
     I_{\vec{\mathbf{V}}} = \frac{1}{d^2_{0}} \sum_{e \in \mathbf{E}} I_e \cdot \left( k_d \cdot G_d (\theta_e, \theta_r)   + k_s \cdot G_s (\theta_e, \theta_m) \right)  + k_a \cdot I_a.
\end{align}
Therein, $G_d (\cdot, \cdot)$ is the importance weight of the screen unit for diffuse reflections, which is given by
\begin{align}\label{eq:impact factor diffuse}
    G_d (\theta_e, \theta_r) = W(\theta_e) \cdot \cos^2 \theta_e \cdot \cos \theta_r.
\end{align}
Moreover, $G_s (\cdot, \cdot)$ is the importance weight of the screen unit for specular reflections, which is expressed as
\begin{align}\label{eq:impact factor specular}
    G_s (\theta_e, \theta_m) = W(\theta_e) \cdot \cos^2 \theta_e \cdot \cos^{n_s} \theta_m.
\end{align}

The above equations tell us: \textit{1) The intensity of the screen light reflected at each face point is a weighted sum of the light from all screen units. 2) The importance weight of each screen unit on the illuminance of each face point is different and mainly depends on the angular variables in the reflection of light.}

\begin{figure}[t]
	\centering
        \includegraphics[width=1.0\linewidth]{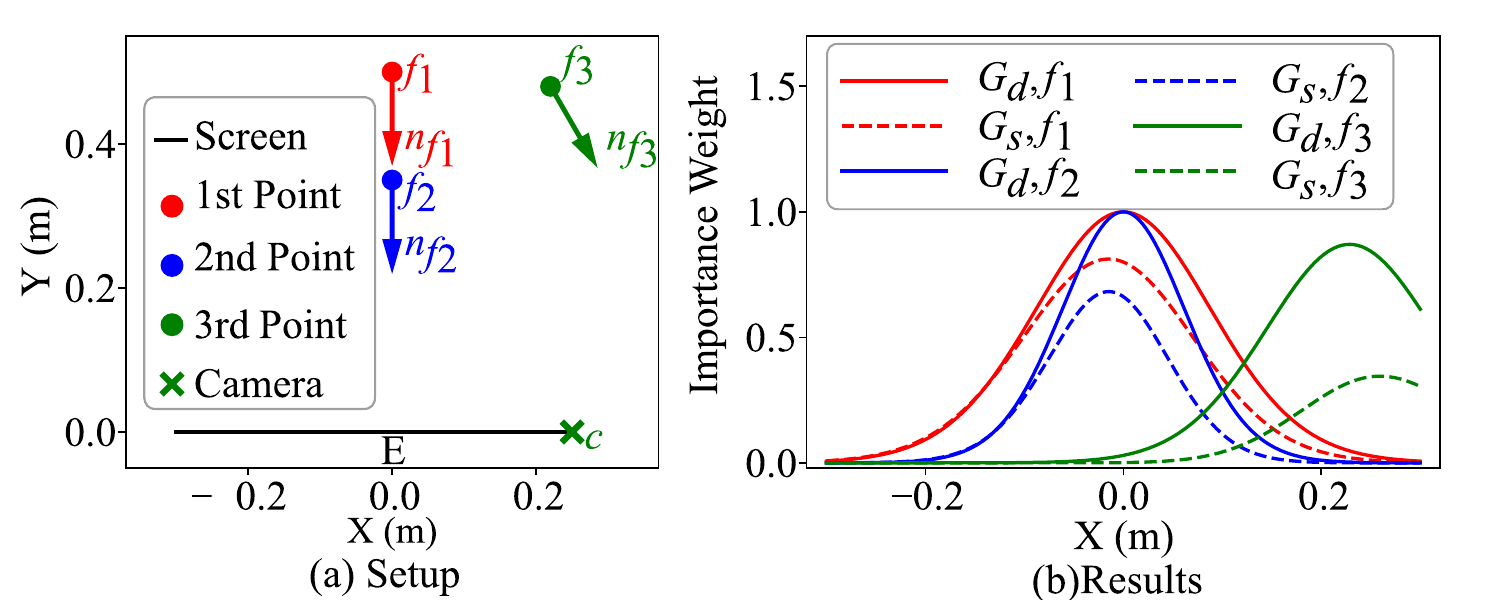}
	\caption{Simulation setup and results.}
	\label{fig:simulation} 
\end{figure}

\subsection{Simulation Validation}

With the above theoretical model, a Python based simulation is conducted to analyze $G_d (\cdot, \cdot)$ and  $G_s (\cdot, \cdot)$ in Eq.~\eqref{eq:impact factor diffuse} and Eq.~\eqref{eq:impact factor specular}.
Specifically, as depicted in Figure~\ref{fig:simulation}~(a), the simulation sets up a 2-dimensional plane and places a screen and a camera on the x-axis. 
Then, three face points with different positions and normal vectors are configured.
In $G_d (\cdot, \cdot)$ and  $G_s (\cdot, \cdot)$, the simulation adopts $W(\theta_e)= \cos^{g} \theta_e $ with $g = 30 $ ~\cite{radiometry2001} and sets the shininess exponent $n_s$ to be 2~\cite{al2011illumination}. 
As shown in Figure~\ref{fig:simulation}~(b), each curve has a peak value in the diffuse reflection, and the peak value is always located around the screen region directly opposite the face point because the angles $\theta_e$ and $\theta_r$ are relatively small in this condition.  
For example, the first face point has relatively high weights on the screen units in the middle area, while the right side of the screen mainly influences the reflection of the third point.
In addition, compared with the first point, the second one has a sharper curve, because it is closer to the screen.
Similar observations can be found in specular reflections.

The above results indicate that: \textit{1) The light reflection off any given face point is not uniformly influenced by the entire screen. 
Instead, it is highly sensitive to a localized "zone" on the screen. 
2) The shape and intensity of this sensitivity zone are unique to the spatial location of one face point and the type of reflection.}

\begin{figure}[t]
	\centering
        \includegraphics[width=1.0\linewidth]{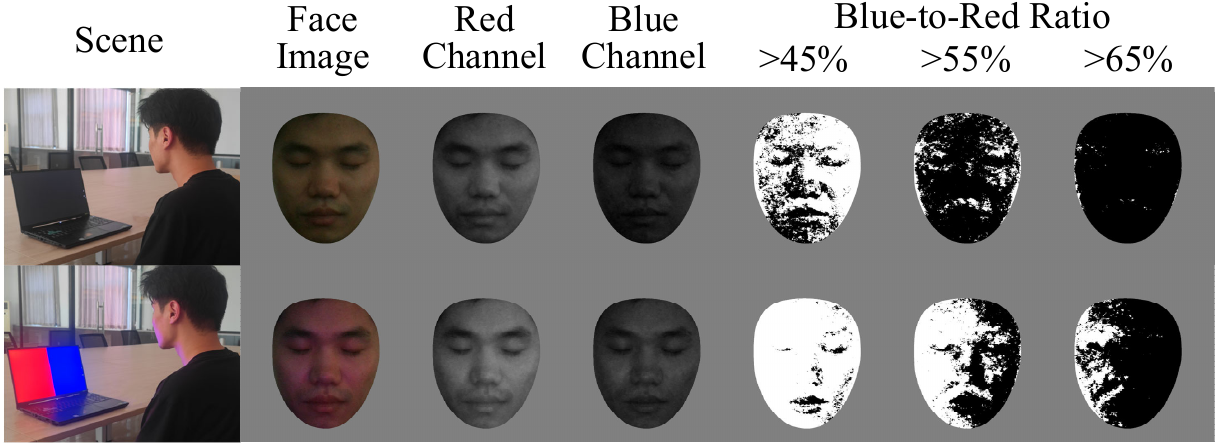}
	\caption{Experimental setup and results.}
	\label{fig:motivtional_experiment} 
\end{figure}

\begin{figure}[t]
	\centering
        \includegraphics[width=1.0\linewidth]{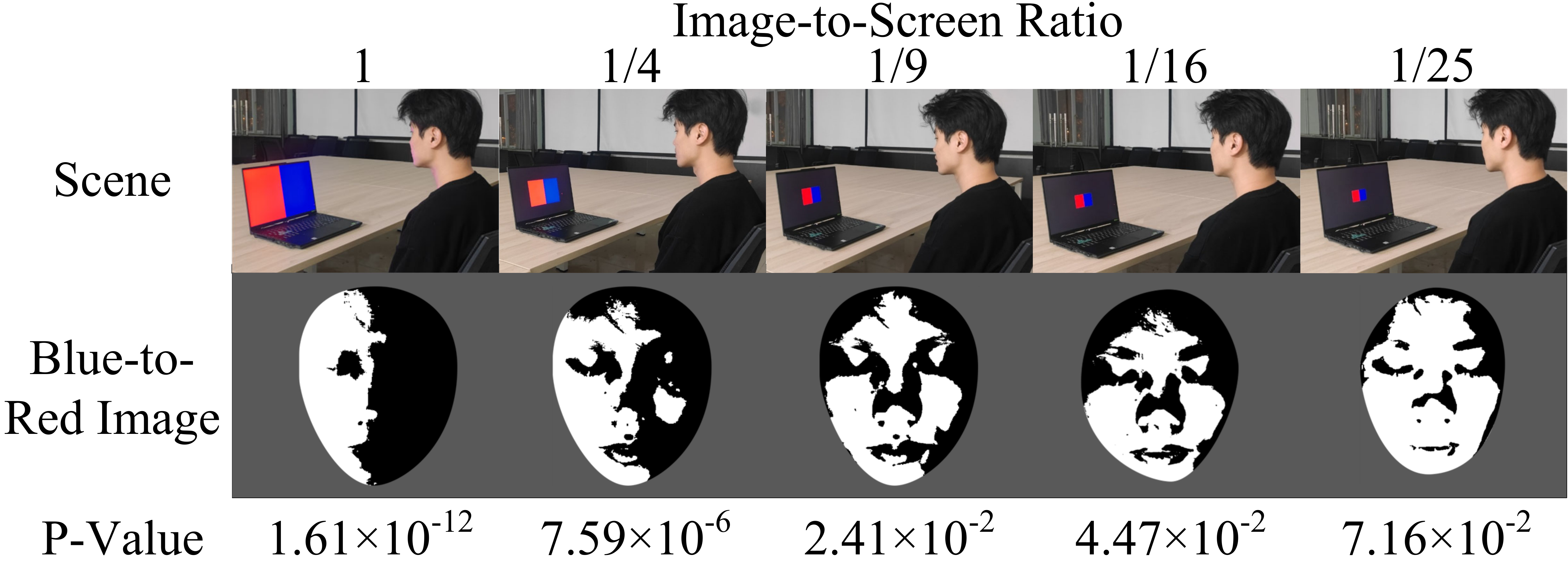}
	\caption{Investigation of minimally differentiable content.}
	\label{fig:minimally_differentiable_content} 
\end{figure}

\subsection{Experimental Validation}
As depicted in Figure~\ref{fig:motivtional_experiment}, one subject sits in front of one laptop and views a black image and a colorful image on the laptop's screen. 
The black image is used to simulate the situation where no light is emitted from the screen.
The left half of the colorful image (1440ps*2560ps) is red, and the right half is blue.
In each case, we leverage the laptop's built-in camera to take a photo of the subject.
After that, we crop the human face out and obtain a clear face image.
As Figure~\ref{fig:motivtional_experiment} shows, when the subject is faced with the colorful image, his face seems more vivid.
To prove this, we extract the red-channel and blue-channel images and observe that they have higher intensity values and become brighter when their faces are in front of the colorful image.

To investigate the impact of different parts of the screen on facial reflections, we show the change in the color composition of each pixel in face images.
For this purpose, we compute a blue-to-red ratio for each pixel by dividing the blue intensity by the red intensity.
Then, we check whether the blue-to-red ratio is bigger than a certain threshold.
If true, the pixel is marked with white; otherwise, it is marked with black.
Three thresholds, i.e., 45\%, 55\%, and 65\%, are selected in each case.
As Figure~\ref{fig:motivtional_experiment} depicts, when the subjects view the black image, blue and red colors are more symmetrically distributed on the entire face.
The reason is that human faces are mainly illuminated by ambient light in this setting, and the right and left sides of the faces have the same illumination condition.
However, when viewing the colorful image, there is blue light distributed on the right-hand side of the subjects' faces and more red light shown on the left-hand side.
This is because the right-hand side of their faces is closer to the right half of the screen, thus reflecting more blue light emitted from that half.

We further investigate the minimally differentiable content (MDC) on the screen (as shown in Figure~\ref{fig:minimally_differentiable_content}), by conducting the Kolmogorov-Smirnov test with a shrunk version of the colorful image.
We observe whether the left and right halves of a blue-to-red image follow the same distribution.
In each shrunk version, we vary the blue-to-red ratio between 1\% and 99\% and select a blue-to-red image with the lowest p-value.
When the shrunk image accounts for 1/16 of the entire screen (360px*640px), the p-value is 0.0416, smaller than a common significance level of 0.05.
When the image-to-screen ratio is 1/25, the p-value is 0.069, larger than 0.05. 
Thus, the MDC of facial reflections is a rectangle occupying 1/16 of the screen area under the experimental settings.

The above experimental results verify that: \textit{1) Different facial regions are sensitive to different screen areas. 2) Facial reflections captured by the webcam can be used to characterize the light patterns of different on-screen contents.}
Thus, facial reflections of an online meeting participant hold the potential to launch screen attacks during video conferencing.

\begin{figure}[t]
    \centering
    \includegraphics[width=\linewidth]{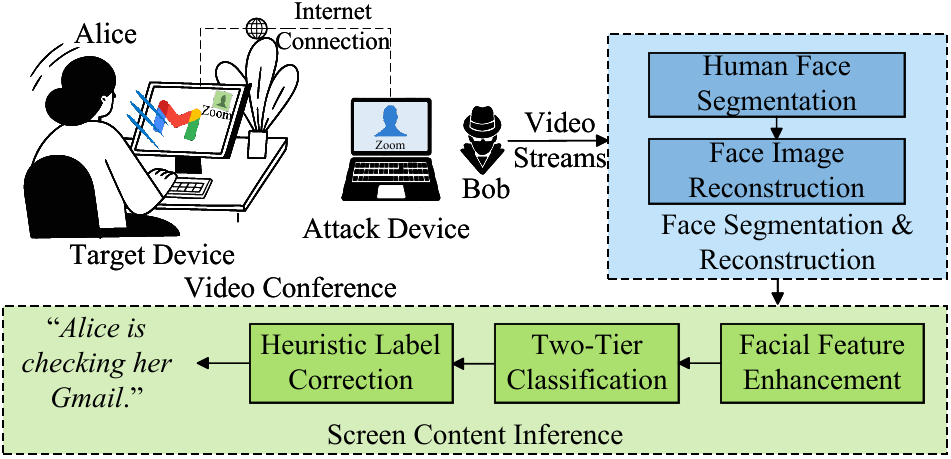}
    \caption{An overview of our FaceTell system.}
    \label{fig-System overview}
\end{figure} 

\section{Design of FaceTell}

\subsection{System Overview}

In this section, we propose FaceTell, a novel screen attack system via pervasive yet unremarkable facial reflections during video conferencing. At a malicious participant, FaceTell can eavesdrop on the victim's multitasking activities without requiring specular objects in the surrounding environment. 
This is achieved by two phases: a training phase and an attacking phase.
In the training phase, FaceTell collects a set of labeled video frames and then trains a dedicated application classifier for screen content inference.
In the attacking phase, FaceTell takes the victim's video frames as input and infers the secondary applications accessed by the victim. 
As shown in Figure~\ref{fig-System overview}, these two phases are implemented by two core components of the FaceTell system: the face segmentation and reconstruction, and the screen content inference. 
We will elaborate on the two components in the following subsections.

\subsection{Face Segmentation and Reconstruction}\label{subsec-segmentaiton-and-reconstruction}

\begin{figure}[t]
	\centering
	\includegraphics[width=1.0\linewidth]{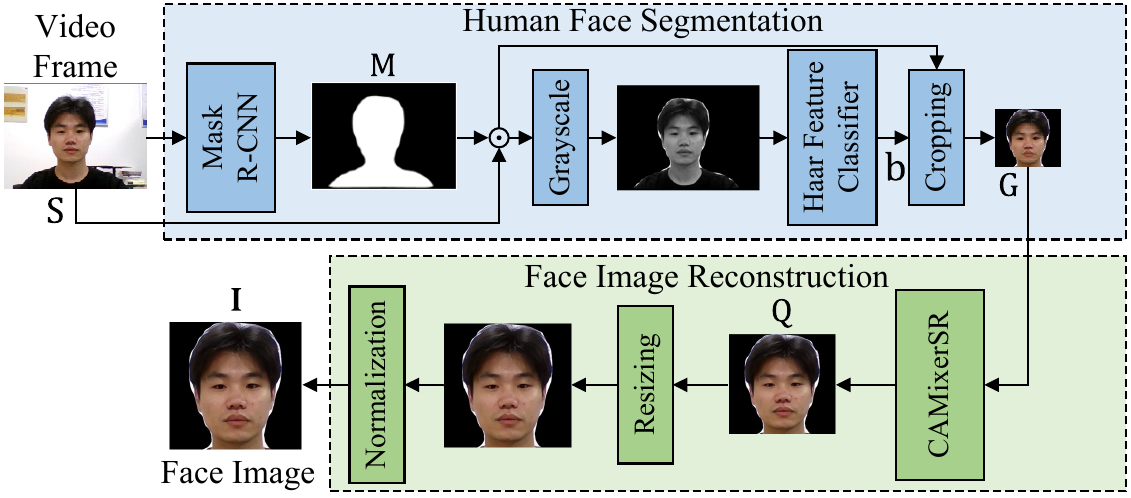}
	\caption{Workflow of face segmentation and reconstruction.}
	\label{fig: face segmentation}
\end{figure}

With the victim's video feed, FaceTell first segments human face images from raw video frames and enhances their quality using super-resolution reconstruction, as shown in Figure~\ref{fig: face segmentation}.

\textbf{Human Face Segmentation.}
In video conferencing, the background within the camera's FoV always changes as meeting environments vary.
More importantly, the virtual background feature is usually enabled when people engage in a video call.
The above reasons render background objects less useful for screen attacks in video conferencing.
In contrast, human faces are generally the foreground objects within the camera's FoV, making them reliable reflectors of light from displays. 
Thus, given a video frame, the first step of FaceTell is to isolate a human face.

To do this, FaceTell samples a video frame from the received video stream every $\delta$ seconds using the OpenCV library.
The reason behind this is that adjacent video frames are very similar due to the temporal continuity of screen content, and sampling video frames can reduce computational complexity without significantly impacting screen content inference.
In our experiment, $\delta$ is set to 0.5 s.
Let us denote $\mathbf{V} \in [0,255]^{ 3 \times W\times H}$ as one extracted RGB frame, where $W$ and $H$ are the width and height.
FaceTell utilizes the mask branch of Mask~R-CNN~\cite{he2017mask}, denoted as $\mathcal{F}_{mask} (\cdot)$, to segment all instances of objects shown in $\mathbf{V}$ and generate a set of masks.
Therein, $\mathbf{M} \in \left\{0,1\right\}^{ W\times H}$ is denoted as a mask matrix for each instance, where $\mathbf{M}(w,h)=1$ indicates that the pixel $(w,h)$ is part of the instance, and $\mathbf{M}(w,h)=0$ indicates the background.
Because surrounding items, such as desks, closets, and appliances, are also recognized by Mask~R-CNN, FaceTell proceeds to pick a human face out of all instances.
It achieves this by exploiting the efficient Haar feature classifier~\cite{viola2001rapid}, which uses a cascade function to identify human faces in images.
Specifically, the Hadamard product is performed between $\mathbf{M}$ and $\mathbf{V}$ to obtain a clear image with an instance only.
Then, the grayscale version of the product result is fed into the Haar feature classifier $\mathcal{F}_{haar} (\cdot)$ for face detection. 
The above process can be expressed as
\begin{align}
    \mathbf{M} & = \mathcal{F}_{mask} (\mathbf{V}), \notag \\
    b & = \mathcal{F}_{haar} ( \text{Gray} (\mathbf{V} \odot \mathbf{M} ) ),
\end{align}
where $\text{Gray} (\cdot) $ is the grayscale function and $\odot$ stands for the Hadamard product.
$b$ is a Boolean value.
If $b$ is true, an instance is detected as a human face.
After that, a clear human face image $\mathbf{G} \in [0,255]^{ 3 \times M \times N }$, having a width of $M$ and a height of $N$, without background objects can be extracted as
\begin{align}
    \mathbf{G} = \text{Crop} (b, \mathbf{V} \odot \mathbf{M}),
\end{align}
where $\text{Crop} (\cdot, \cdot)$ is the cropping operation that generates the bounding box and crops a face image out when $b$ is true.

\begin{figure}[t]
    \centering
    \begin{subfigure}[b]{0.48\linewidth}
        \centering
        \includegraphics[width=\textwidth]{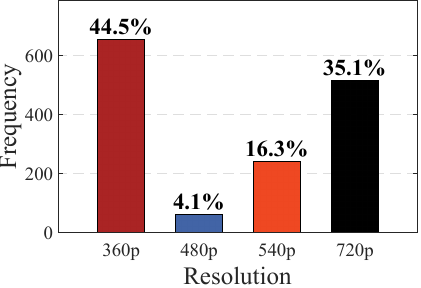}
        \caption{Histogram of video resolutions.}
        \label{fig:resolution}
    \end{subfigure}
    \hfill
    \begin{subfigure}[b]{0.48\linewidth}
        \centering
        \includegraphics[width=\textwidth]{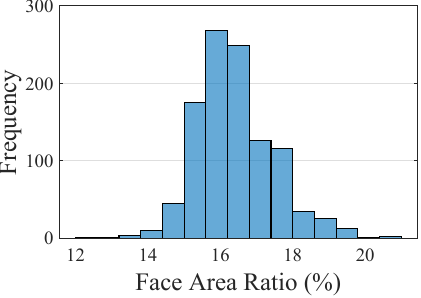}
        \caption{Histogram of face area ratios in raw video frames.}
        \label{fig:face_area}
    \end{subfigure}

    \caption{Characteristics of online video streams on Zoom.}
    \label{fig:online_video_characteristics}
\end{figure}

\textbf{Face Image Reconstruction.}
Due to limitations in camera capability and variable network bandwidth, the video resolution during an online meeting is low and variable.
To verify this, we collect an online video stream lasting 25 minutes from Zoom, extract one frame per second, and record the frame resolution.
As depicted in Figure~\ref{fig:online_video_characteristics}~(a), the video resolution of online meetings is dynamic, ranging from 360p to 720p, and more than 40\% of the extracted frames have a resolution of 360p, which is very low compared to high-end telescopic lens photos~\cite{backes2008compromising,backes2009tempest}. 
What makes things worse is that a human face is only a small part of a video frame.
We crop 1,000 face images using the above face segmentation method and compute their area ratios in the raw video frames.
As reported in Figure~\ref{fig:online_video_characteristics}~(b), human faces occupy about 16\% of pixels in a video frame on average.
The above two reasons render it challenging to extract effective information directly using the cropped face images.

To alleviate this problem, FaceTell transforms the low-quality face images into high-resolution versions using the CAMixerSR network~\cite{wang2024camixersr}, an image super-resolution technique that is based on a content-aware mixer that assigns convolution for simple contexts and additional deformable window attention for sparse textures.
Taking the face image $\mathbf{G}$ as input, the CAMixerSR network $\mathcal{F}_{sr} (\cdot)$ generates a higher-quality image $\mathbf{Q} \in [0,255]^{3 \times 2M\times 2N}$ by upscaling the height and the width of $\mathbf{G}$ at two times as $\mathbf{Q} = \mathcal{F}_{sr} (\mathbf{G})$.
Because the size of the upscaled image $\mathbf{Q}$ is changing among different frames, FaceTell resizes $\mathbf{Q}$ into a square image with a fixed size of $L\times L$.
In addition, the resized image is further normalized using a z-score normalization within each channel for effective hidden feature extraction.
The final face image $\mathbf{I} \in [0,255]^{3 \times L \times L}$ can be obtained by $\mathbf{I} = \text{Norm} ( \text{Resize} (L,\mathbf{Q}) )$,
where $\text{Resize} (\cdot)$ and $\text{Norm} (\cdot)$ represent the resize and normalization operations, respectively. 

\subsection{Screen Content Inference}

\begin{figure}[t]
	\centering
	\includegraphics[width=1.0\linewidth]{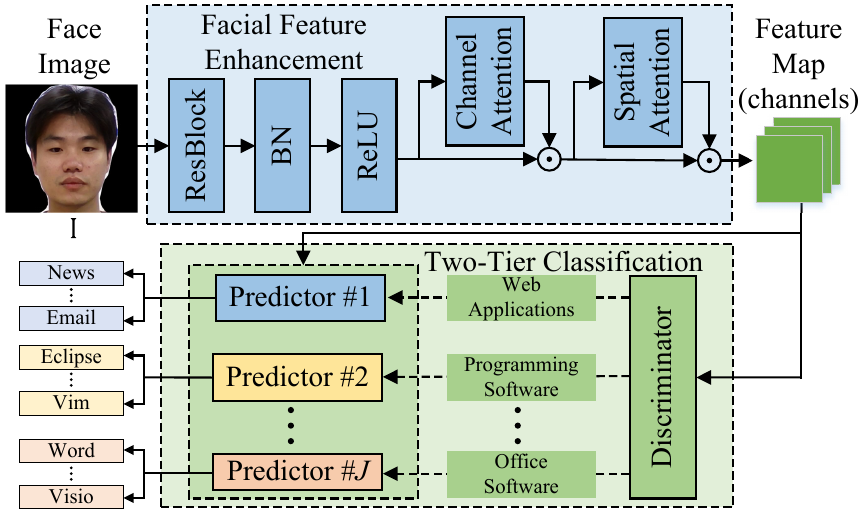}
	\caption{The framework of FaceTell's screen content inference component.} 
	\label{fig: feature extraction and classification}
\end{figure}

As depicted in Figure~\ref{fig: feature extraction and classification}, the second step of FaceTell is to perform screen content inference based on face images.

\textbf{Facial Feature Enhancement.} 
In this component, FaceTell takes a face image $\mathbf{I}$ as input and generates a high-level feature map.
It first leverages a residual convolutional block, which includes a main path and a shortcut connection path, to extract content-related features for effective image feature abstraction.
Three convolutional kernels with a size of $3 \times 3 $ are used in the main path.
Their stride and padding size are both set to 1.
A 2-dimensional batch normalization (BN) layer and a rectified linear unit (ReLU) are followed for feature representation and non-linearity introduction.
In this way, an intermediate feature map $\mathbf{C} \in \mathbb{ R }^{ 3 \times L \times L} $ is obtained.

Next, FaceTell exploits a convolutional block attention module~\cite{woo2018cbam} (CBAM) to further enhance its feature representation capability.
The CBAM comprises a channel attention submodule and a spatial attention submodule, which are used to capture significant features in the channel and spatial dimensions, respectively. 
Therein, the channel attention submodule has a max-pooling (MaxPool) branch and an average-pooling (AvgPool) branch.
Two branches are followed by a shared multi-layer perceptron (MLP).
Their intermediate features are aggregated and fed into a sigmoid function, consequently yielding a channel attention map.
When it comes to the spatial attention submodule, it applies average-pooling and max-pooling operations along the channel axis and concatenates their outputs, which are further processed by a convolution layer (Conv) with a kernel size of $7 \times 7 $ and a sigmoid function to generate a spatial attention map.
In the CBAM, the intermediate feature map $\mathbf{C}$ is first fed into the channel attention submodule and multiplied with the channel attention map using the Hadamard product.
Then, the product result $\mathbf{C'} \in\mathbb{ R }^{ 3 \times L \times L} $ is processed by the spatial attention submodule in the same way to obtain the enhanced feature map.
Mathematically, the final feature map $\mathbf{S} \in\mathbb{ R }^{ 3 \times L \times L} $ after the CBAM can be expressed as
\begin{align}
    \mathbf{C'} = \phi (\text{MLP} (\text{MaxPool} (\mathbf{C}) ) +  \text{MLP} ( \text{AvgPool} (\mathbf{C}))) \odot \mathbf{C}, \notag \\
        \quad \mathbf{S}  =  \phi ( \text{Conv} (\text{MaxPool} (\mathbf{C'}) \oplus  \text{AvgPool} (\mathbf{C'})) ) \odot \mathbf{C'},
\end{align}
where $\phi (\cdot)$ denotes the sigmoid function and $ \oplus$ is the concatenation operation.

\textbf{Two-Tier Classification.}
Despite the enhanced feature map, it is still challenging to accurately recognize fine-grained screen content due to the large number of online applications and software. 
FaceTell addresses this challenge by devising a two-tier classification model to first classify screen content into several main categories, i.e., web applications, programming software, office software, etc., and then predict the finer-grained applications in this category. 
For example, the web application category includes reading news, online shopping, and checking emails. 
The two-tier classification model achieves this by employing a category discriminator and a set of application predictors.

Specifically, the category discriminator transforms each feature map $\mathbf{S}$ into a category label $\mathbf{g} \in \left\lbrace 0,1 \right\rbrace^{J}$, which is the one-hot encoding vector that indicates the category of applications shown on the screen.
Therein, $J$ is the category number.
The category discriminator consists of three fully connected (FC) layers with 512, 256, and $J$ neurons, respectively, followed by a softmax function to yield a probability vector $\mathbf{p} \in [0,1 ]^{J}$. 
Then, FaceTell encodes the probability vector $\mathbf{p}$ into the one-hot vector $\mathbf{g}$ corresponding to the $j$-th category.

After determining the category of screen content, the application predictor belonging to the $j$-th category is selected and facilitates fine-grained application classification.
The selected predictor transforms each feature map $\mathbf{S}$ into a probability vector $\mathbf{q} \in [0,1 ]^{K_j}$.
$K_j$ is the number of computer applications in the $j$-th category.
For simplicity, each application predictor has the same network architecture as the category discriminator.
Similarly, the probability vector $\mathbf{q}$ is converted into an application label $\mathbf{r} \in \left\lbrace 0,1 \right\rbrace^{K_j}$, which is the one-hot encoding vector that indicates which application the victim is using.

For each face image, we combine its category-application label pair of $\mathbf{g}$ and $\mathbf{r}$ into a unified label $ \mathbf{y} \in \left\lbrace 0,1 \right\rbrace^{K} $, where $\sum_{j=1}^{J}K_j = K$, to represent a specific application.
Given a conferencing video stream, FaceTell can generate a content label sequence as $\mathcal{Y} = \left\lbrace  \mathbf{y}^{1}, \cdots, \mathbf{y}^{t}, \cdots, \mathbf{y}^{T}  \right\rbrace$, where $\mathbf{y}^{t}$ is the estimated label at time $t$ and $T$ is the sequence length.

\begin{algorithm}[t]
    \caption{Heuristic Label Correction}\label{algo: label correction}
    \begin{algorithmic}[1]
        \Require Predicted label sequence $\mathcal{Y}$ and its length $T$
        \Ensure Corrected label sequence $\mathcal{Z}$
        \State $t \Leftarrow 1 $
        \If{$ t \le T \: \: \text{and} \: \: \text{Start}(t) = \text{True}$} \: \: \: \:  // a step starts
         \State $\mathbf{z}^t \Leftarrow \mathbf{y}^t $ \: \: \: \: // known content
        \State $t \Leftarrow t + 1 $
        \If{$\text{End}(t) = \text{False}$} \: \: \: \:  // a step continues
        \State $\mathbf{z}^t \Leftarrow \mathbf{z}^{t-1} $
        \State \textbf{goto} \text{line 4}
        \Else \: \: \: \: //a step ends
        \State \textbf{goto} \text{line 2}
        \EndIf
        \Else
        \State $\mathbf{z}^t \Leftarrow (0, \cdots , 0) $ \: \: \: \: // unknown content
        \State $t \Leftarrow t + 1 $
        \State \textbf{goto} \text{line 2}
        \EndIf
    \end{algorithmic}
\end{algorithm}

\textbf{Heuristic Label Correction.}
In this component, FaceTell takes a further step to correct face images that are mistakenly classified.
Misclassifications could stem from either inadequate features extracted from a single face image or screen content unseen in the training phase.
To deal with this issue, FaceTell proposes a novel heuristic label correction (HLC) algorithm.
The correction algorithm is based on the observation that it generally takes a while for users to interact with a computer application, and users are not likely to switch them too frequently.
For example, in~\cite{firstpagesage2022session}, it reports that web sessions typically last 2 to 3 minutes on average.
Thus, we have prior knowledge that, in most cases, the current prediction label estimated by our predictor is likely to be the same as the previous one. 
Additionally, face images whose labels appear frequently within a period are likely to be correctly classified.

Based on the above observation, a corrected content label sequence can be considered as a step function, which is a piecewise constant function having finite steps.
Let us denote the corrected label sequence as $ \mathcal{Z} = \left\lbrace \mathbf{z}^{1}, \cdots,\mathbf{z}^{t}, \cdots, \mathbf{z}^{T},  \right\rbrace$, where $ \mathbf{z} \in \left\lbrace 0,1 \right\rbrace^{K} $ could be one of $K$ one-hot vectors for predicted applications or an all-zero vector for other unknown content.
Using the above notations, we define two functions regarding the two label sequences $\mathcal{Y}$ and $\mathcal{Z}$ as follows.

\textit{Definition 1 (Start of A Step): } At time $t$, let us denote $\mathbf{W}_t =  \left\lbrace \mathbf{y}^{t}, \mathbf{y}^{t+1}, \cdots, \mathbf{y}^{t+T_s-1}  \right\rbrace $ as a segment of the label sequence $\mathcal{Y}$ with a length of $T_s$. 
We can determine that a step of $\mathcal{Z}$ starts at the time $t$ or not as
\begin{align}\label{eq: step start}
    \text{Start} (t) =  \left\lbrace 
    \begin{array}{lcl}
    \text{True} \: , \:  \text{if} \: \: \frac{\text{Count}(\mathbf{y}^{t}, \mathbf{W}_t)}{T_s} \ge  \sigma_s ; \\
    \text{False} \: , \: \text{else} .
\end{array}
\right.
\end{align}
Therein, $0.5 \leq \sigma_s<1$ is a determining threshold and $\text{Count}(\mathbf{y}^{t},\mathbf{W}_t)$ represents the count number of the label $\mathbf{y}^{t}$ present in the segment $\mathbf{W}_t$.
This definition means that when the current label $\mathbf{y}^{t}$ accounts for the majority of the next $T_s$ labels in the sequence $\mathcal{Y}$, the start of a step in $\mathcal{Z}$ begins.

\textit{Definition 2 (End of A Step): } At time $t$, we denote $V_{t}(\tau) = \left\lbrace \mathbf{y}^{t}, \cdots, \mathbf{y}^{t+\tau}  \right\rbrace$, where $\tau \ge 0$.
It represents a segment of the label sequence $\mathcal{Y}$ with a varying length of $\tau+1$.
An end of a step of $\mathcal{Z}$ is detected or not as
\begin{align}
    \text{End} (t) =  \left\lbrace 
    \begin{array}{lcl}
    \text{False} \: , \:  \text{if} \: \: \exists \: 0 \le \tau \le T_e, \text{s. t.,} \: \frac{\text{Count}(\mathbf{z}^{t-1}, V_{t}(\tau))}{\tau+1} \ge  \sigma_e ; \\
    \text{True} \: , \: \text{else},
\end{array}
\right.
\end{align}
where $T_e$ stands for the maximum window size of $V_{t}(\tau)$ and $0<\sigma_e  \leq 0.5 $ is another determining threshold.
$\text{Count}(\mathbf{z}^{t-1}, V_{t}(\tau))$ represents the count number of the previous step label $\mathbf{z}^{t-1}$ present in the segment $V_{t}(\tau)$.
This definition indicates that when the previous label $\mathbf{z}^{t-1}$ never has a high proportion in any segment of the next $T_e$ labels in $\mathcal{Y}$, the end of a step in $\mathcal{Z}$ is recognized.

Based on the defined functions $\text{Start}(\cdot)$ and $\text{End}(\cdot)$, a heuristic label correction algorithm is proposed in Algorithm~\ref{algo: label correction}.
Its core idea is to alternately use $\text{Start}(\cdot)$ and $\text{End}(\cdot)$ on the raw label sequence $\mathcal{Y}$ to estimate the time interval of a label step and convert all involving labels in $\mathcal{Z}$ into the same.
An illustration of our algorithm is depicted in Figure~\ref{fig: correction_algorithm}.
At the beginning, the function $\text{Start}(\cdot)$ is used to detect the start of a step.
If a step starts, we continuously leverage the function $\text{End}(\cdot)$ to identify the end of a step.
If the end is identified, the function $\text{Start}(\cdot)$ is called again to find the start of the next step.
Moreover, if a predicted label does not belong to any step, we set it to an all-zero vector, indicating unknown content.
As demonstrated in Figure~\ref{fig: correction_algorithm}, the 5th, 8th, 12th, 13th, 16th, and 23rd predicted labels are corrected by our algorithm.
In our experiments, we empirically set \(\sigma_s=0.90\), \(T_s=10\), \(\sigma_e=0.10\), and \(T_e=10\).
Towards this end, our heuristic label correction algorithm $\mathcal{F}_{correction} (\cdot)$ outputs a corrected label sequence as $\mathcal{Z} = \mathcal{F}_{correction} (\mathcal{Y}).$

\begin{figure}[t]\label{fig: correction_algorithm}
	\centering
\includegraphics[width=1.0\linewidth]{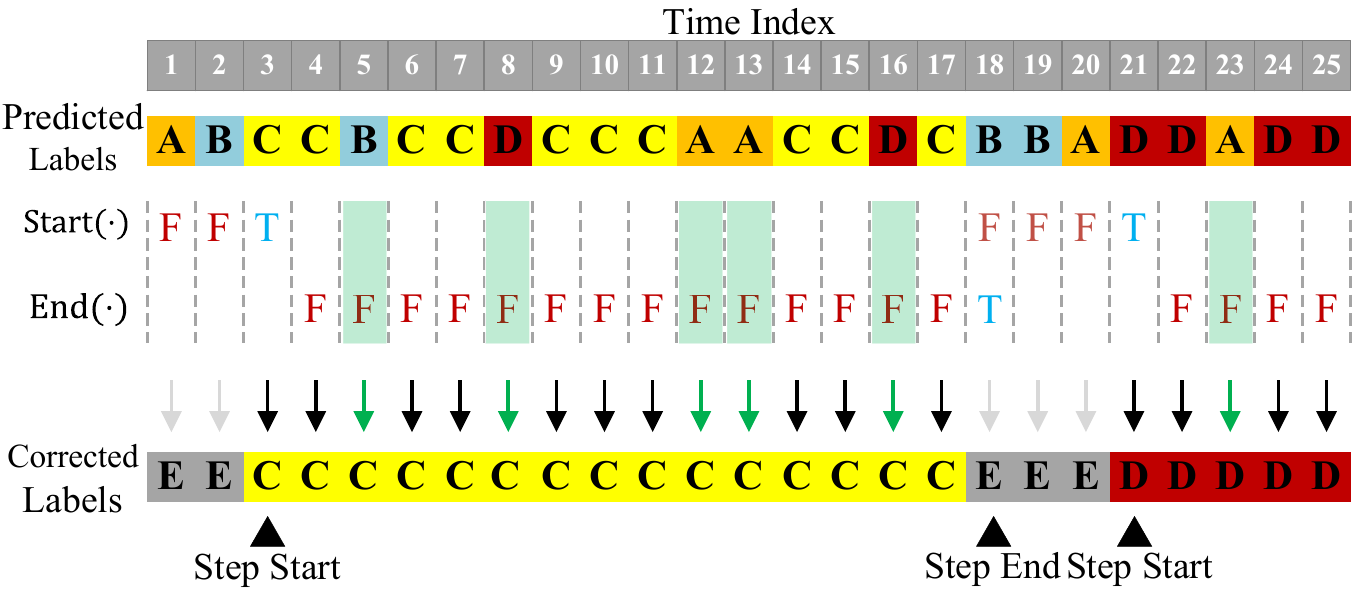}
	\caption{Illustration of heuristic label correction. A, B, C, and D stand for four application labels, and E is for unknown ones. T and F denote True and False, respectively. In this setting, \(\sigma_s=0.6\), \(T_s=8\), \(\sigma_e=0.4\), and \(T_e=6\). A green rectangle means the occurrence of a label correction.}\label{fig: correction_algorithm}
\end{figure}

\begin{table} 
\caption{Screen content of interest.}
\label{tab: screen content}
\small
\setlength{\tabcolsep}{3pt}
\begin{tabular}{cclll}
\cline{1-2}
Category             & Application                                                                                                      & \multicolumn{1}{c}{} & \multicolumn{1}{c}{} & \multicolumn{1}{c}{} \\ \cline{1-2}
Web Applications      & \begin{tabular}[c]{@{}c@{}}Searching, News, Shopping, \\ Email, Social, Sports\end{tabular}               & \multicolumn{1}{c}{} & \multicolumn{1}{c}{} & \multicolumn{1}{c}{} \\
Office Software      & \begin{tabular}[c]{@{}c@{}}Word, Excel, Visio, OneNote,\\ PowerPoint, Adobe Reader,\end{tabular}          & \multicolumn{1}{c}{} & \multicolumn{1}{c}{} & \multicolumn{1}{c}{} \\
Programming Software & \begin{tabular}[c]{@{}c@{}}MATLAB, Visual Studio, Eclipse,\\ PyCharm, Vim, Notepad\end{tabular}            & \multicolumn{1}{c}{} & \multicolumn{1}{c}{} & \multicolumn{1}{c}{} \\
Multimedia Software          & \begin{tabular}[c]{@{}c@{}}Tik Tok, Disney, Dts Sound, Pandora,\\ Hulu, Netflix, iHeartRadio, Crunchyroll\end{tabular}                    &                      &                      &                      \\
OS Applications       & Files Searching, OS Setting                                                                                 &                      &                      &                      \\
Video Conferencing   & \textbackslash{}                                                                                          &                      &                      &                      \\ \cline{1-2}
\end{tabular}
\end{table}

\begin{figure}[t]
    \centering
    \includegraphics[width=1.0\linewidth]{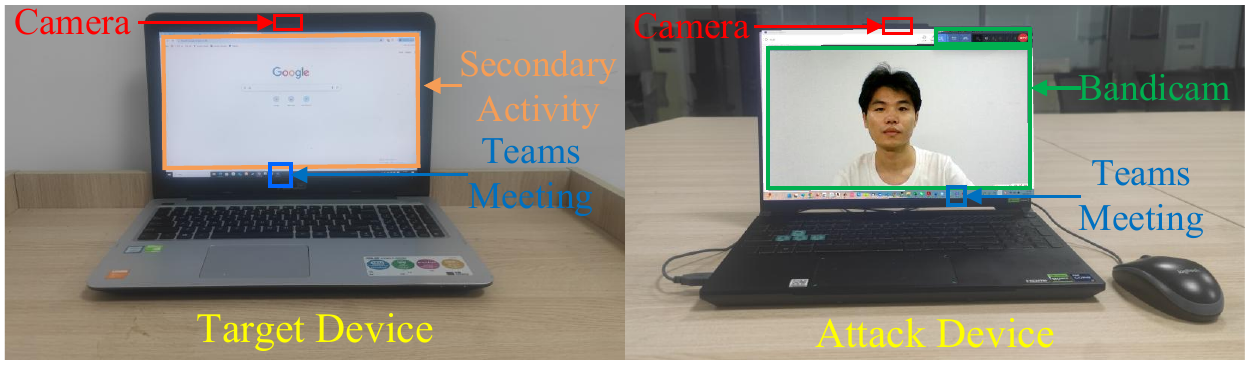}
    \caption{Device setups during data collection.}
    \label{fig-layout}
\end{figure} 

\begin{figure*}[t]
    \centering
    \includegraphics[width=\textwidth]{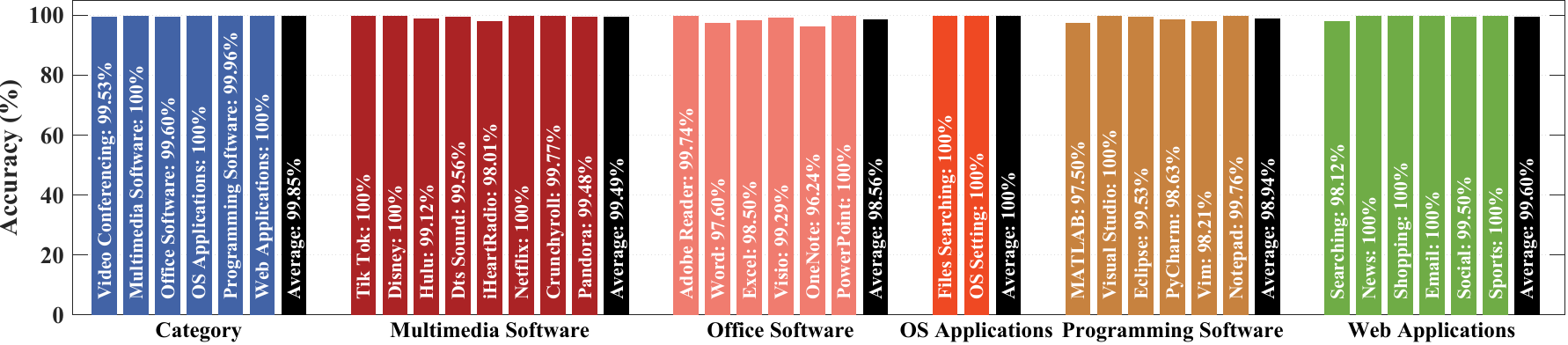}
    \caption[width=\textwidth]{Overall performance of FaceTell.}
    \label{fig-Classification Accuracy}
\end{figure*}

\section{Implementation and Setup}\label{sec-system_implementation}


\textbf{Target and Attack Devices.} 
We use four consumer-grade laptops as target and attack devices in our experiments.
Specifically, three of them serve as target devices, responsible for capturing videos of victims.
The specifications of the three target devices are as follows: the ASUS TUF A16 is equipped with a USB 2.0 HD UVC WebCam; the Lenovo Y7000P has a 720p HD camera; the HP ProBook 440 is equipped with a 720p HD camera.
In addition, a Lenovo Y9000P with a 720p HD camera is used as an attack device to record video streams from the other devices.
Note that all devices have an auto-brightness-adjustment feature, which is activated by Windows automatically.

\textbf{Video Conferencing Platforms.} 
We select four mainstream applications that enable video calls: Skype, Zoom, Teams, and WeChat, and install them on the target and attack devices.
These applications have a diverse range of underlying video encoding/decoding protocols and network transmission optimization strategies.

\textbf{Screen Content of Interest.} 
As shown in Table~\ref{tab: screen content}, we select 28 representative computer applications when people multitask in online meetings.
They are divided into five categories, including web applications, office software, programming software, multimedia software, and OS applications.
Moreover, we add another category of video conferencing, which indicates that an online meeting participant does not conduct secondary activities.


\textbf{Data Collection.} 
We recruit 24 subjects, including 14 males and 10 females, to collect online video streams in 13 different environments, containing 3 bedrooms, 7 offices, and 3 dormitories.
All subjects are college students, aged from 21 to 25 years old.
Some of them wear eyeglasses or face masks.
With the 24 subjects, we collect video streams in 24 rounds, resulting in more than 12 hours of video data.
In each round, the attack device and one target device are placed in two different rooms and set to join the same online meeting with their cameras activated.
One subject is invited to sit within the FoV of the target device to act as a victim.
Initially, the subject does not engage in any secondary activity for two minutes.
The subject then minimizes the video conference window and sequentially uses each application listed in Table~\ref{tab: screen content} for one minute.
We record the usage time of each application for sample annotation. 
At the same time, the attack device runs Bandicam to capture video streams from the target device, as illustrated in Figure~\ref{fig-layout}.
In this way, a 30-minute video is collected in each round.

\textbf{Datasets.} 
We convert each collected video into a group of image frames with a sampling rate of 2~Hz. 
Using the proposed face segmentation and reconstruction approach in~\cref{subsec-segmentaiton-and-reconstruction}, a clear face image is obtained with a size of $800 \times 800 $ for each frame.
We sort images within each group based on time stamps and annotate them with corresponding category and application labels.
In this way, we obtain two disjoint datasets, i.e., a training dataset and a testing dataset.
Specifically, the training dataset contains 49,407 face images, and the testing dataset has 12,372 face images.

\textbf{Training Details.}
We implement FaceTell on an OMNISK AIX7550-G3 workstation running on Ubuntu 20.04.6TLS with 32GB RAM and 4 NVIDIA GeForce RTX3090 graphics cards.
The category discriminator and application predictors are built using PyTorch and optimized using the cross-entropy loss on the training dataset.
The Adam optimizer with a learning rate of $1 \times 10^{-4}$ and a batch size of 16 is adopted.
The model is optimized over five epochs.

\textbf{Evaluation Metric.}
We adopt accuracy as the main evaluation metric.
It is defined as the ratio of correctly classified samples to the total number of samples as
\begin{equation}
    \text{Accuracy} = \frac{\text{Number of Correct Predictions}}{\text{Total Number of Samples}}.\notag
    \label{eq:accuracy}
\end{equation}

\begin{table*}[t]
    \centering 
    \caption{Ablation study on FaceTell.} 
    \label{tab:ablation_results} 
    
    \definecolor{Gray}{gray}{0.9}
    \small 
    \setlength{\tabcolsep}{2.5pt}
    \begin{tabular}{l|cccccc}
        \toprule
        \textbf{Variants} & \textbf{Category} & \textbf{\thead{Multimedia\\Software}} & \textbf{\thead{Office\\Software}} & \textbf{\thead{OS\\Applications}} & \textbf{\thead{Programming\\Software}} & \textbf{\thead{Web\\Applications}} \\
        \midrule
        \#1: w/o CAMixerSR and HLC        & 89.88\% & 82.40\% & 82.81\% & 90.06\% & 83.56\% & 82.72\% \\
        \#2: w/o Channel Attention and HLC & 90.78\% & 81.49\% & 82.65\% & 89.24\% & 82.37\% & 81.23\% \\
        \#3: w/o Spatial Attention and HLC & 91.38\% & 81.99\% & 83.49\% & 90.76\% & 82.91\% & 82.43\% \\
        \#4: w/o Two-Tier Classification and HLC & \textbackslash & 74.14\% & 73.09\% & 72.66\% & 72.87\% & 77.38\% \\
        \#5: w/o HLC                     & 92.56\% & 84.44\% & 84.98\% & 91.70\% & 85.09\% & 84.22\% \\
        \midrule
       
        \textbf{FaceTell (Complete Model)} & \textbf{99.85\%} & \textbf{99.49\%} & \textbf{98.56\%} & \textbf{100\%} & \textbf{98.94\%} & \textbf{99.60\%} \\
        \bottomrule
    \end{tabular}
\end{table*}

\begin{figure}[t]
    \centering
    \includegraphics[width=\linewidth]{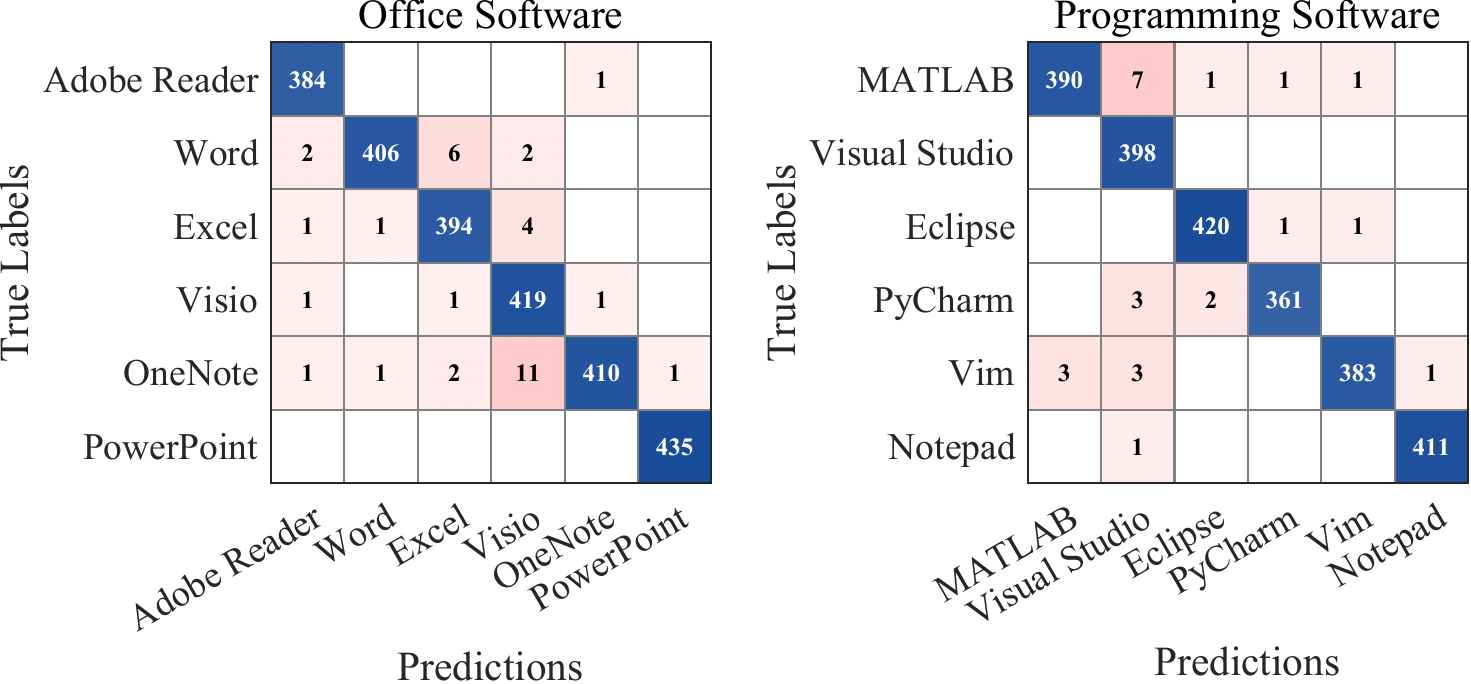}
    \caption{Confusion matrices for office software and programming software.}
    \label{fig:confusion_matrix}
\end{figure} 

\section{Evaluation Results}

\subsection{Overall Performance}\label{sec-overall_performance}
We first present the overall performance of FaceTell.
To do this, we evaluate our system on the testing dataset and report system performance for coarse-grained content category prediction and fine-grained application prediction in Figure~\ref{fig-Classification Accuracy}.
As the figure shows, FaceTell achieves an average accuracy of 99.85\% for category prediction.
In particular, it obtains a prediction accuracy of 100\% in multimedia software, OS applications, and web applications.
This is because multimedia software and web applications generally have a more colorful layout, which is significantly different from other applications.
As for OS applications, the screen would display a large amount of blank area in the file searching, and mostly shows the system's default theme color in the OS setting, making them more distinguishable.
When it comes to fine-grained application prediction, FaceTell still presents high performance in each category with an average accuracy of 99.32\%.
It has a relatively low accuracy of 98.56\% in office software and an accuracy of 98.94\% in programming software.
To understand the underlying reasons, we plot their confusion matrices in Figure~\ref{fig:confusion_matrix}.
We find that there are some misclassifications among Word, Excel, Visio, and OneNote in the office software.
This is because these applications belong to the Microsoft Office Suite and share many similarities in UI layouts, such as the ribbon interface and quick access toolbar, to provide users with a consistent experience.
As for programming software, FaceTell yields some wrong predictions on MATLAB, Visual Studio, Eclipse, PyCharm, and Vim.
The reason may be that these applications generally follow a conventional integrated development environment (IDE) layout with a menu bar at the top, a main editor window, and sidebar panels.
Despite that, FaceTell still achieves a prediction accuracy of more than 98\% in the two categories.

\subsection{Ablation Study}

We proceed to conduct an ablation study to validate the effectiveness of each core component within FaceTell. 
For this purpose, we create five variants by systematically ablating specific components. 
The variants are described below.
1) The first variant removes the CAMixerSR network and the HLC algorithm.
2) The second variant ablates the channel attention submodule in the CBAM block, and the HLC algorithm is also omitted.
3) The third variant discards the spatial attention submodule in the CBAM block and also omits the HLC algorithm.
4) The fourth variant exploits an application classifier to directly recognize the 28 applications without two-tier classification and the HLC algorithm.
5) The fifth variant removes only the HLC algorithm.
As Table~\ref{tab:ablation_results} shows, the accuracy of each variant is at least 7\% lower than that of FaceTell in each task, suggesting the effectiveness of each component proposed in FaceTell.
Specifically, the first three variants perform lower than the fifth, which indicates that the lack of face image reconstruction or facial feature enhancement will lead to insufficient feature representation of facial reflections and a decrease in accuracy.
Moreover, the proposed two-tier classification model is also important to FaceTell.
Without it, the fourth variant suffers an accuracy decrease of at least 6.84\% when compared to the fifth variant.
We can also observe that the label correction algorithm plays an important role in FaceTell and leads to an accuracy increase from 7.29\% to 27.34\%.
The above observation proves that the temporal continuity of human-computer interactions is useful in correcting mistakenly predicted labels.

\begin{figure}[t]
    \centering
    \includegraphics[width=\linewidth]{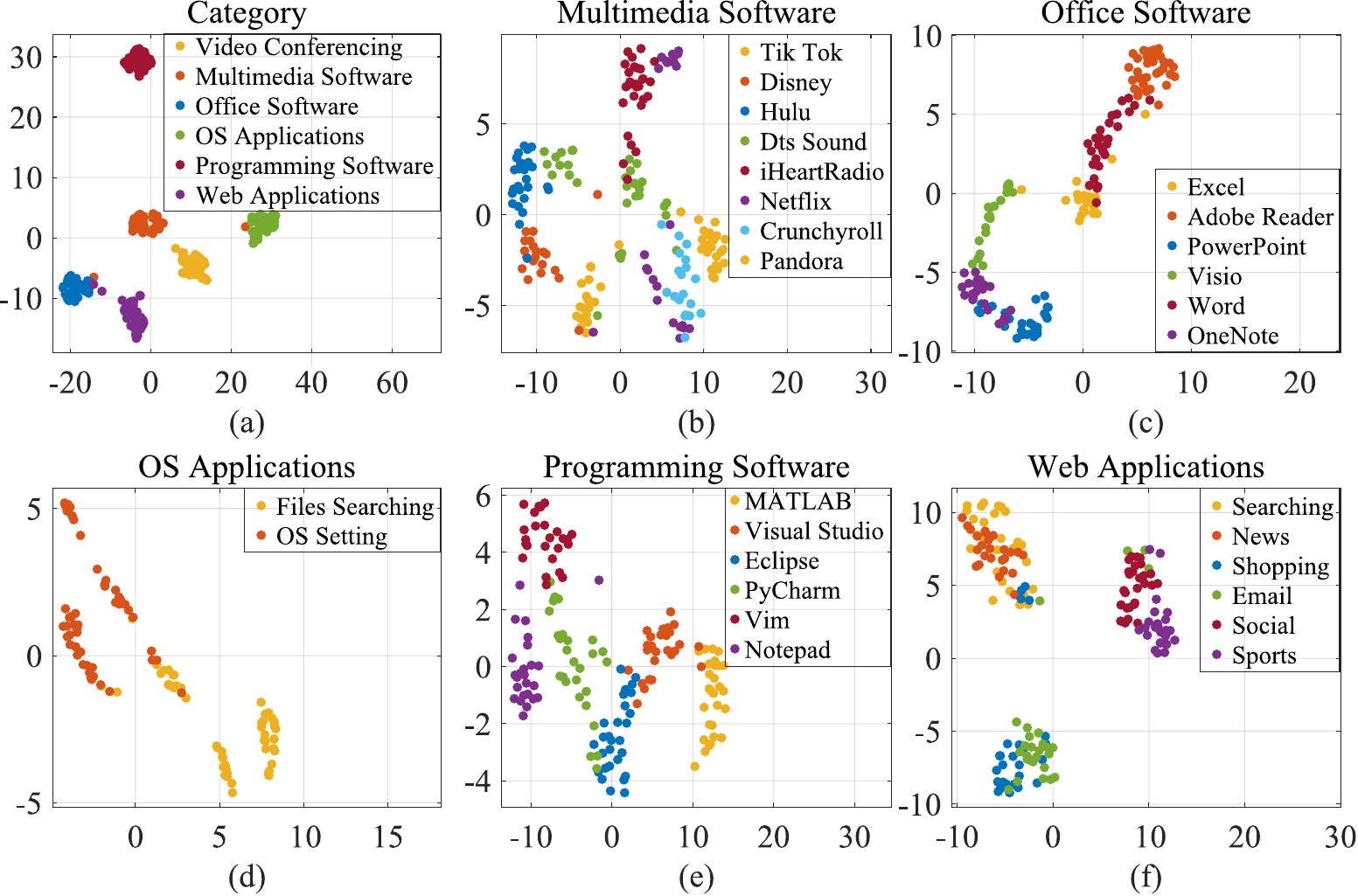}
    \caption{Feature visualization of the category discriminator and application predictors using t-SNE.}
    \label{fig-tsne}
\end{figure} 

\subsection{Feature Visualization}
We exploit the t-distributed stochastic neighbor embedding (t-SNE) technique to visualize the output of the second FC layer of the category discriminator and five application predictors, respectively.
In this way, six scatter plots are present in Figure~\ref{fig-tsne}, where each point corresponds to a labeled sample.
As Figure~\ref{fig-tsne}~(a) shows, the category discriminator presents a good separability of different kinds of samples.
Six groups of points can be clearly observed.
In each group, feature points are highly compact.
Moreover, each group is distinct from the others, except for a few outliers.
In addition, the points in the video conferencing group are the farthest from those of the others, which tells us that the light variations of video conferencing are quite different from those of the secondary activities.
The results indicate that each category has a unique pattern of facial reflections.
As for the application predictors, the feature points within different groups are relatively scattered.
Some points belonging to different groups are even intermixed, but the overall boundaries can still be roughly observed.
This is because the applications within the same category probably have similar UI layouts and generate indistinguishable facial reflections in some cases.

\subsection{Time Consumption and Scalability}
We report FaceTell's runtime on the OMNISK AIX7550-G3 workstation.
We extract 1,000 video frames from the collected video, feed them into FaceTell, and record the average runtime of the five core components, i.e., human face segmentation, face image reconstruction, facial feature enhancement, two-tier classification, and heuristic label correction, per sample.
Compared to the first three components, the two-tier classification and heuristic label correction have a negligible runtime of less than 6 $\mu$s due to their low computational complexity.
Moreover, the face image reconstruction has a runtime of 102.50 ms, because the super-resolution operation is relatively time-consuming.
Nevertheless, FaceTell has an average runtime of approximately 124 ms per frame.
To further demonstrate its scalability with multiple victims, we build a real-time version of FaceTell on a computer with a 62.4 GB memory CPU and a 24 GB memory GPU.
With only Zoom activated, the CPU (GPU) usage is 10.4\% (0\%). 
If eavesdropping on one Zoom user, FaceTell increases the CPU (GPU) usage to 20.9\% (62\%). 
When eavesdropping on two victims simultaneously, the CPU (GPU) usage is 26.8\% (98\%). 
With these hardware configurations, FaceTell can attack two victims simultaneously.

\subsection{Impact Factors}\label{subsec-impact_factor}

In this subsection, we investigate FaceTell's performance under different impact factors, including subject gender, facial occlusions, distances, angles, and ambient light.

\begin{figure}[t]
    \centering
    \includegraphics[width=\linewidth]{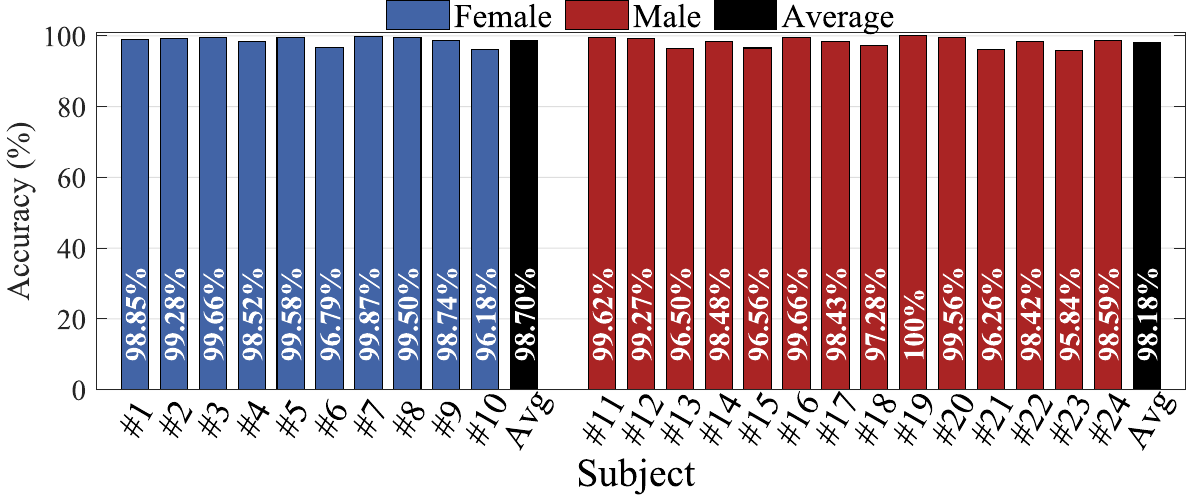}
    \caption{Performance on different subjects.}
    \label{fig-user_accuracy}
\end{figure}

\textbf{Impact of Subject Gender.} 
There are 24 subjects, including 10 women and 14 men, involved in our experiments.
The subject's gender may have an impact on FaceTell, as women and men exhibit different facial features in terms of skin texture, hair characteristics, etc.
To understand this impact, we report the average accuracy of the application prediction for all subjects in Figure~\ref{fig-user_accuracy}.
FaceTell achieves an accuracy of at least 96.18\% in the female group and an accuracy of more than 95.84\% in the male group.
When it comes to overall performance, the average accuracy for female subjects is 98.70\%, and that for male subjects is 98.18\%, indicating a minimal performance difference between women and men.
The above results indicate that the subject's gender has a negligible impact on FaceTell.

\begin{table}[t]
\centering
\caption{Performance under different occlusion conditions.}
\label{tab:occlusion_accuracy}
\setlength{\tabcolsep}{3pt} 
\begin{tabular}{lcc}
\toprule
\textbf{Occlusion} & \textbf{Accuracy w/o HLC} & \textbf{Accuracy w/ HLC} \\
\midrule
No Occlusion & 93.31\% & 99.43\% \\
Eyeglasses   & 92.62\% & 99.93\% \\
Face Masks         & 90.95\% & 99.64\% \\
\bottomrule
\end{tabular}
\end{table}

\begin{figure}[t]
    \centering
    \includegraphics[width=\linewidth]{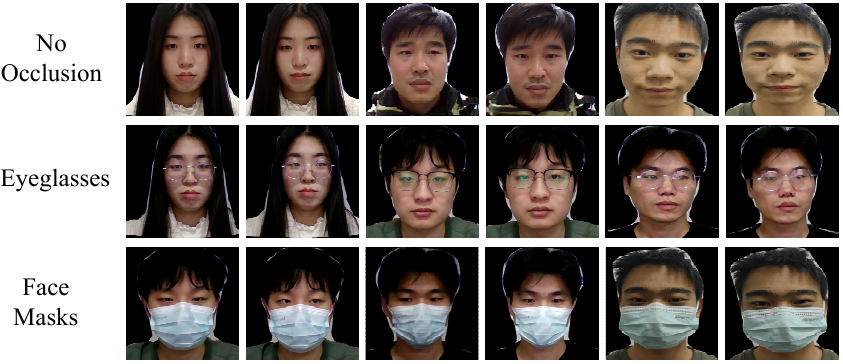}
    \caption{Face images under different occlusion conditions.}
    \label{fig-occlusion}
\end{figure} 

\textbf{Impact of Facial Occlusions.} 
In practice, facial occlusions may appear during video conferencing and change the facial reflections of participants.
To explore this impact, we compute the average accuracy under various facial appearance conditions, including no occlusion, wearing eyeglasses, and wearing face masks, with and without the heuristic label correction.
The results are reported in Table~\ref{tab:occlusion_accuracy}.
Before the label correction, FaceTell achieves the best performance with an accuracy of 93.31\% when no occlusion is present, and has the lowest accuracy of 90.95\% in the scenario of wearing face masks.
To understand the underlying reasons, we showcase face images under the three types of facial occlusions in Figure~\ref{fig-occlusion}.
When no occlusion is shown on the subjects' faces, there are many highlighted areas, such as foreheads, cheeks, and noses, that can generate obvious facial reflections. 
However, when subjects wear eyeglasses, the lenses do not always show recognizable screen content.
Since specular reflections have strict requirements on the angles between the screen, the camera, and the lenses, the images on subjects' eyeglasses sometimes come from the items outside of the screen and weaken the light illumination from the screen, resulting in lower performance.
As for face masks, they cover most of the area around the mouth and reduce the area of facial reflections. 
As a result, the performance of FaceTell is the worst in this case.
Despite that, the prediction accuracy of FaceTell exceeds 99\% in each scenario after the label correction.
We can conclude that FaceTell is robust to different types of facial occlusions in real-world video conferencing scenarios.

\begin{figure}[t]
    \centering
    \includegraphics[width=\linewidth]{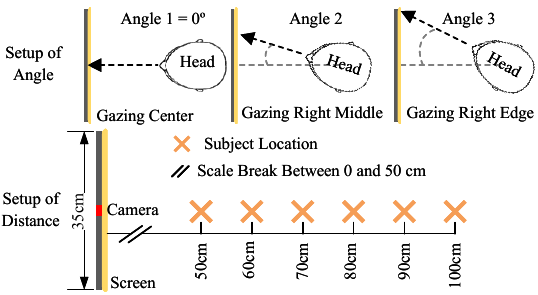}
    \caption{Setups of different distances and angles.}
    \label{fig-distance_angle}
\end{figure} 

\textbf{Impact of Distance and Angle.}
As indicated in Eq.~\eqref{eq:revised reflected inrensity}, the distance and angle between a face point and the screen determine the intensity of the screen light and would impact FaceTell's performance.
Because it is non-trivial to investigate the impact of distance and angle on a face point basis in practice, we choose to study the impact of overall distance and angle between a subject's face and the screen.
To do this, we collect online video streams under different distance and angle settings.
As illustrated in Figure~\ref{fig-distance_angle}, one subject is recruited to sit at a fixed location and gaze at the center of the screen, the middle of the right half of the screen, and the right edge of the screen to simulate three angles, i.e., Angle 1, Angle 2, and Angle 3, between the whole face and the screen. 
In addition, we vary the distance between the screen and the subject location from 50 cm to 100 cm.
The setting ranges of angles and distances conform to people's daily computer usage habits in practice. 
In each setting, a 30-minute video is collected and fed into FaceTell for screen content inference.
The application prediction performance with and without the HLC algorithm is illustrated in Table~\ref{tab:accuracy_compact}.
In each angle setting, the accuracy stays stable at first and then gradually declines after 70 cm without the label correction.
This is because the increase in distance will weaken the intensity of facial reflections and reduce the proportion of a face in the camera's FoV, rendering FaceTell harder to capture light emanations from the screen.
Moreover, in each distance setting, as the angle increases, the accuracy of FaceTell declines.
This is because the increase in angle will reduce the face area that can reflect screen light.
Despite these, the label correction can significantly improve the accuracy to more than 96.81\% in all cases.
The above results show the robustness of FaceTell to practical variations in the face-screen distances and angles.

\begin{table}[t]
\centering
\caption{Performance under different distances and angles.}
\label{tab:accuracy_compact}
\footnotesize
\setlength{\tabcolsep}{2pt} 
\begin{tabular}{
    S[table-format=3.0]  
    S[table-format=2.2]  
    S[table-format=2.2]  
    S[table-format=2.2]  
    S[table-format=2.2]  
    S[table-format=2.2]  
    S[table-format=2.2]  
}
\toprule
{\textbf{Distance}} &
\multicolumn{2}{c}{\textbf{Angle 1}} & \multicolumn{2}{c}{\textbf{Angle 2}} & \multicolumn{2}{c}{\textbf{Angle 3}} \\
\cmidrule(lr){2-3} \cmidrule(lr){4-5} \cmidrule(lr){6-7}
{(cm)} &  
{\textbf{w/o HLC}} & {\textbf{w/ HLC}} & {\textbf{w/o HLC}} & {\textbf{w/ HLC}} & {\textbf{w/o HLC}} & {\textbf{w/ HLC}} \\

\midrule
50  & 91.49 & 99.76 & 91.33 & 99.76 & 90.31 & 97.39 \\
60  & 92.74 & 98.70 & 91.40 & 99.63 & 90.86 & 97.79 \\
70  & 92.75 & 99.42 & 92.41 & 98.64 & 91.18 & 98.43 \\
80  & 91.87 & 97.24 & 90.92 & 97.79 & 89.08 & 99.35 \\
90  & 89.54 & 98.99 & 88.89 & 97.08 & 86.97 & 97.56 \\
100  & 86.01 & 96.81 & 85.15 & 97.23  & 83.36 & 97.76 \\
\bottomrule
\end{tabular}
\end{table}

\begin{figure}[t]
    \centering
    \includegraphics[width=\linewidth]{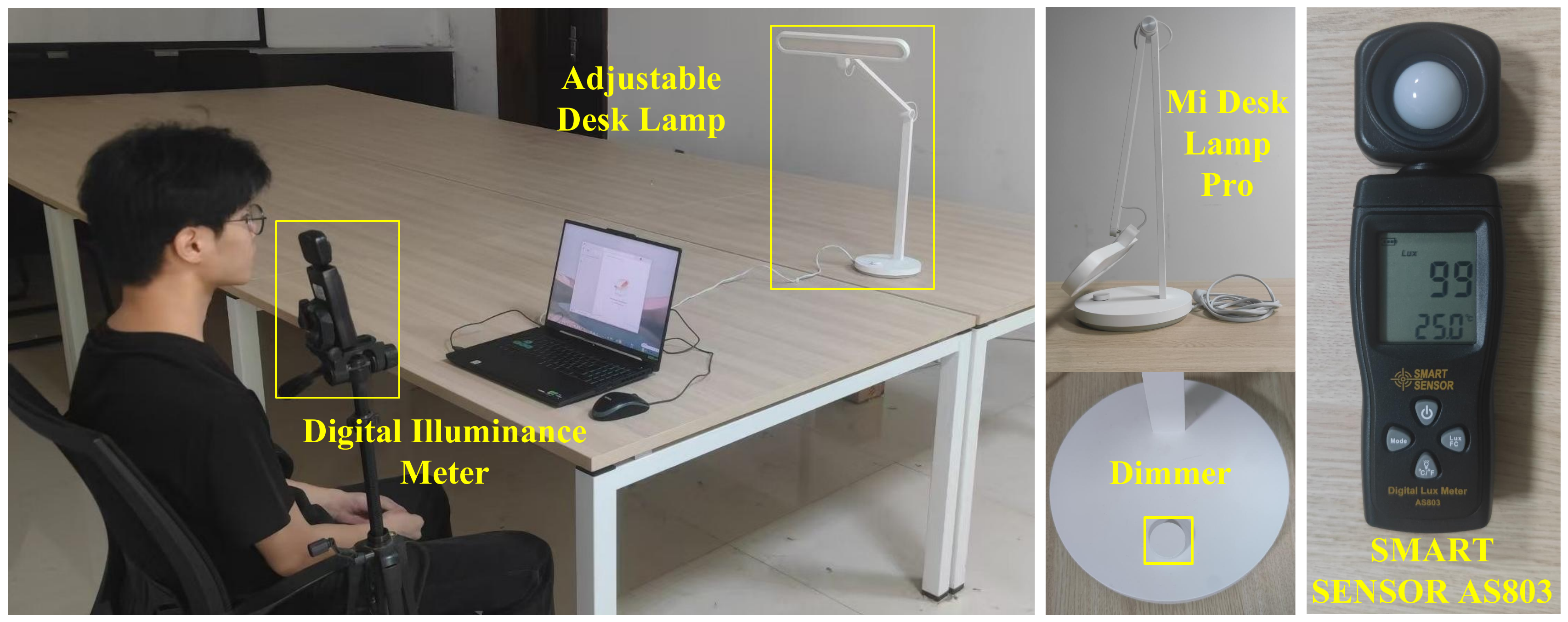}
    \caption{Testbed for investigation of ambient light.}
    \label{fig-ambient_setup}
\end{figure} 

\begin{figure}[t]
    \centering
    \includegraphics[width=\linewidth]{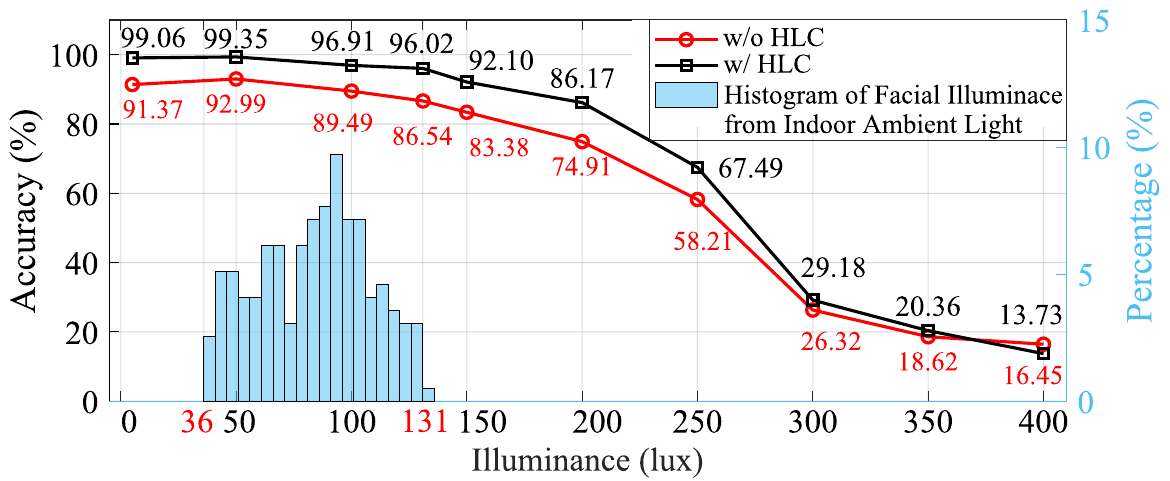}
    \caption{Performance under different intensities of ambient light.}
    \label{fig-illumination}
\end{figure} 

\textbf{Impact of Ambient Light.}
As depicted in Figure~\ref{fig-ambient_setup}, we build a testbed with a SMART SENSOR AS803 illuminance meter and an adjustable Mi Desk Lamp Pro to collect video data under different intensities of ambient light.
First, we use the illuminance meter to measure light illuminance in 5 rooms, including a meeting room, a corridor, a laboratory, a dormitory, and a classroom, under normal lighting conditions.
In each room, the illuminance meter is held vertically to simulate the light reception of human faces and record 20 measurements in lux.
Then, we plot the histogram of the collected illuminance measurements in Figure~\ref{fig-illumination} and find that facial illuminance ranges from 36 to 131~lux in indoor environments. 
Then, a subject is asked to sit in front of the ASUS laptop in an office with closed curtains and turned-off lights. 
The adjustable lamp is placed 1 meter away from the laptop to serve as the source of ambient light, and the illuminance meter is vertically mounted near the subject's face and towards the screen.
Using this testbed, we simulate facial illuminance of 5, 50, 100, 131, 150, 200, 250, 300, 350, and 400 lux from ambient light. 
In each illuminance setting, the laptop's screen is first turned off, and the desk lamp is adjusted to generate a specified illuminance. 
Figure~\ref{fig-illumination} shows that, as the illuminance of ambient light increases, FaceTell's performance degrades.
This is because facial reflections are a composition of ambient light and screen light as indicated in Eq.~\eqref{eq:revised reflected inrensity}.
Thus, the increase in ambient light will reduce the proportion of screen light in facial reflections, making FaceTell more difficult to extract light variations on the subject's face.
However, in the indoor facial illuminance range, FaceTell still achieves a high accuracy from 96\% to 99\%.
The above results demonstrate that FaceTell has resilience to ambient light in indoor environments.

\subsection{Practicality Analysis}
According to the experimental results in~\cref{subsec-impact_factor}, the performance of FaceTell is significantly influenced by face-screen distance and ambient light intensity. 
More specifically, as shown in Table~\ref{tab:accuracy_compact} and Figure~\ref{fig-illumination}, the results indicate that FaceTell can achieve $\ge 96\%$ accuracy when a victim is 50 cm to 100 cm away from the screen and exposed to an ambient facial illuminance $< 130$ lux. 
We believe that these settings conform to practical computer usage and indoor illumination. First, the concluded face-screen distance range is in line with the user-screen distance (i.e., 25 inches/63.5 cm) recommended by the American Academy of Ophthalmology to minimize eye strain and prevent visual fatigue~\cite{daniel2025digital}.
Second, considering that the vertical-to-horizontal illuminance ratio in typical office environments ranges from 0.3 to 0.5 ~\cite{dias2017toward}, the ambient facial illuminance of 130 lux corresponds to approximately 260 to 430 lux on indoor work planes.
This illuminance range includes the recommended light levels of many typical indoor environments, such as homes, coffee break rooms, and classrooms~\cite{toolbox2004illuminace}.   

\subsection{Failure Cases}
Although FaceTell can obtain high performance in most cases, we report the conditions where our system is limited.

\textbf{Excessive Face-Screen Distances.}
We increase the face-screen distances to 120 cm and 150 cm and test the accuracy of FaceTell. The accuracy drops to 75\% and 25.65\%, respectively. In practice, such excessive face-screen distances may not be realistic as they cause visual discomfort (e.g., eye strain, blurry vision).

\textbf{High Ambient Light Intensities.}
As shown in Figure 18, when the facial illuminance from ambient light exceeds 300 lux, FaceTell’s accuracy drops to below 30\%. This is because the high ambient light overwhelms the facial optical variations caused by the screen. 
However, such high-intensity frontal lighting is practically undesirable, as excessive vertical illuminance causes discomfort glare and visual fatigue ~\cite{zhang2022study}, leading users to naturally avoid these conditions.

\textbf{New Applications.}
For the new applications that are not listed in Table~\ref{tab: screen content}, FaceTell may experience poor performance.
However, if a new application belongs to a defined category in Table~\ref{tab: screen content}, FaceTell can recognize its category, and hence guarantee the basic prediction performance for monitoring user activities. 
To verify this, samples from a new application, YouTube, are collected for testing. 
The discriminator classifies these samples as multimedia software with an accuracy of 82.88\%. 
For those applications that do not belong to any categories in Table~\ref{tab: screen content}, the adaptation only requires fine-tuning the discriminator and predictors, without necessitating a complete retraining of the entire model from scratch. 

\textbf{Dark Mode Settings.}
Dark mode is a user interface display setting that employs a color scheme with light-colored text and icons on a dark background. By design, this mode significantly reduces the overall light intensity emitted from the screen, thus hampering FaceTell's performance. 
To verify this, we collect new data from OS applications under the dark mode setting and test the category discriminator on them. 
The accuracy of category prediction decreases to 64.13\%, which, however, is still much higher than the random guess baseline of 16.67\% (1/6).
One possible solution to combat this issue is to retrain FaceTell using video data from different color settings of applications.  

\section{Related Work}


\textbf{Screen Attacks.} 
The foundational work in this domain is established by Backes et al.~\cite{backes2008compromising}.
By exploiting optical reflections off objects such as teapots, eyeglasses, and even eyeballs, they recover content from an LC screen from a distance of over 30 meters using a telescope.
In their subsequent work~\cite{backes2009tempest}, they improve the quality of images recovered from eyeball reflections to infer on-screen information. 
Using consumer-grade cameras, Raguram et al.~\cite{raguram2011ispy} achieve automatic reconstruction of typed input by capturing the key pop-out effect of virtual keyboards of smartphones in reflections off sunglasses. 
The work~\cite{xu2013seeing} tracks the motion trajectories of a user's fingers from eyeball reflections to predict human-device interactions. 

With the proliferation of video conferencing, the attack vector has shifted from close-proximity observations to online video feeds. 
Zachary et al.~\cite{weinberg2011still} exploit the background area of a webcam’s FoV to detect uniform-color images on the screen. 
However, the background area can be easily blurred by the virtual background feature. 
Moreover, this approach is susceptible to human movements. Instead, we choose to exploit indispensable human faces, the foreground area of the FoV.
However, the foreground part is dynamic and much smaller than the background part. 
To address this challenge, FaceTell employs Mask R-CNN and a Haar feature classifier for human face detection and segmentation. 
Then, the super-resolution CAMixerSR network is applied to enhance the quality of a segmented face image. 
Both the prior works~\cite{long2023private,wasswa2022proof} leverage the foreground eyeglasses as attack surfaces.
Private Eye~\cite{long2023private} investigates the physical limits of reconstructing textual information from eyeglass reflections in video conferencing. 
Hassan et al.~\cite{wasswa2022proof} leverage the power of image processing and computer vision techniques to predict the kind of content displayed on the victim’s screen. 
However, eyeglasses cannot serve as a reliable source of compromising reflections, because not everyone wears eyeglasses, and eyeglass reflections do not always come from screens. 
To transcend this limitation, FaceTell exploits non-glossy human faces as attack surfaces. 
Using human faces is generally more challenging because facial reflections are a mixture of specular and diffuse reflections, making it harder to recognize the screen contents. 
To tackle this challenge, we devise a dedicated neural network, comprising a CBAM block and an effective two-tier classification model, for content label estimation and propose an HLC algorithm to boost our prediction performance.

\textbf{Other Side-Channel Attacks.}
Synesthesia~\cite{Daniel2019acoustic} exploits acoustic noise emitted by an LC screen to eavesdrop on screen content in a voice call remotely.
Similarly, Cecconello et al.~\cite{cecconello2019skype} recover device input by analyzing keyboard acoustic emanations transmitted during a Skype call.
Homma et al.~\cite{hayashi2014EM} leverage electromagnetic (EM) leakage from a tablet and reconstruct the screen image from 2 meters away.
Ni et al.~\cite{ni2023fingerprints} recover fingerprint images from in-display fingerprint sensors in smart devices via EM signals.
WaveSpy~\cite{li2020wavespy} probes the state of LC molecules on a screen using the mmWave radar, allowing the eavesdropping of screen content through the wall. 
AppListener~\cite{ni2023applistener} infers mobile app activities by analyzing RF energy harvested from Wi-Fi signals.
Fang et al.~\cite{fang2018csi} exploit disturbances in Wi-Fi signals to infer keyboard input. 
This paper exploits a new side channel, i.e., facial reflections, to eavesdrop on the screen activities of online meeting participants.


\section{Limitations}\label{sec: limitations}

We conduct further experiments to evaluate FaceTell in three challenging scenarios: 1) using additional screens, 2) operating in outdoor sunny environments, and 3) applying skin smoothing filtering. 
The results are reported in Table~\ref{tab:limitations}.

\begin{table}[t] 
\centering
\caption{FaceTell's performance under challenging scenarios.}
\label{tab:limitations}
\begin{tabular}{lc}
\toprule
\textbf{Challenging Scenario} & \textbf{Accuracy} \\
\midrule
Additional Screens & 15.09\% \\
Outdoor Environments & 13.99\% \\
Skin Smoothing Filters & 17.91\% \\
\bottomrule
\end{tabular}
\end{table}

\textbf{Using Additional Screens.} 
The attack model of FaceTell has an implicit assumption: the primary dynamic illumination on the victim's face originates from a single screen.
However, a video conferencing participant may have multiple screens, e.g., a programmer may have a multi-monitor setup to improve work efficiency. 
To investigate FaceTell performance under such interference, we test it using data collected from the victim who faces two screens. The accuracy of the FaceTell attack is 15.09\%.
The reason is that the additional light source produces overlapping illumination on the victim's face. 
Consequently, changes in facial reflections cannot be accurately attributed to a specific screen's activity, resulting in some misclassifications or unknown categories.
We intend to explore this direction in our future work.

\textbf{Operating in Outdoor Sunny Environments.}
We extend our evaluation to outdoor environments. We collect facial videos outdoors, where a subject sits in a shady spot on a sunny day (4:20 pm to 5:00 pm). 
The environmental illuminance is approximately 1300 lux. 
In this condition, an accuracy of 13.99\% is achieved, suggesting that meeting outdoors can effectively withstand ``FaceTell'' attacks.

\textbf{Applying Skin Smoothing Filters.}
During a video conference, a participant may use skin smoothing filters.  
These filters apply a smoothing effect to the facial regions in video frames. 
This results in difficulty for FaceTell to obtain application-related optical features. 
To verify this, FaceTell is tested when the skin smoothing filter is enabled on Zoom.
It has a low accuracy of 17.91\%.

We hope that by presenting the above limitations of FaceTell, we can help other researchers understand future research directions on screen attacks.
Meanwhile, to maximize this possibility, we have open-sourced the code to facilitate subsequent researchers to reproduce our system.

\section{Countermeasures}~\label{subsec-countermeasures}

\textbf{Introducing Additional Dynamic Light Sources.}
As discussed in~\cref{sec: limitations}, additional dynamic light sources will degrade FaceTell's performance.
This strategy constitutes a form of active interference at the physical level. 
A user can place a secondary dynamic light source, e.g., another turned-on display, in front of his face but outside the camera's FoV.
This source generates a task-irrelevant facial illumination and prevents the accurate activity inference of FaceTell.

\textbf{Enabling Skin Smoothing Filters.} 
As suggested in~\cref{sec: limitations}, enabling skin smoothing filters can significantly worsen FaceTell's performance and thus is an effective countermeasure to human face-based screen attacks. 
However, these filters will make a participant's face look less realistic, causing some participants to be reluctant to enable them to preserve an authentic communication experience.

\textbf{Adding Adversarial Perturbations.}
To generate effective and imperceptible software-level filters, adversarial perturbations are a promising solution.
They bring minimal changes to input images and can steer a classifier away from true predictions~\cite{goodfellow2014explaining}.
Thus, dedicated adversarial perturbations can be generated based on our discriminator and predictors and added to facial regions before sending video feeds to other online meeting participants.
In this way, the perturbed video frames are received by FaceTell and would cause wrong predictions regarding the secondary activities of other participants.

\section{Conclusion}
In this paper, we present a new side channel to launch screen attacks using facial reflections in video conferencing.
To validate the feasibility, we propose FaceTell, the first automated system that exploits pervasive yet subtle facial reflections to eavesdrop on the secondary activities of online meeting participants.
Experimental results show that FaceTell has a high application prediction accuracy of 99.32\% for 28 computer applications.
Also, FaceTell has an average runtime of about 124~ms for each inference.
We also evaluate the resilience of FaceTell to different practical impact factors, including subject gender, facial occlusions, distances, angles, and ambient light.
Potential countermeasures are discussed to mitigate the FaceTell attack.
We expect that our research results can increase public awareness and stimulate deeper research on this side-channel attack, as well as new defense strategies. 

\section*{Acknowledgments}

This work was supported in part by the National Natural Science Foundation
of China with Grant 62301499 and the Henan Association for Science and Technology with Grant 2025HYTP037.
Wanqing Tu’s work is supported by the EPSRC and DSIT through the Communications Hub CHEDDAR (grant numbers EP/X040518/1 and EP/Y037421/1), and the European Union’s Horizon Europe research and innovation programme grant “SENSORBEES” (agreement no. 101130325). 




\cleardoublepage
\appendix
\section*{Ethical Considerations}

Our research explores a novel side-channel attack vector in video conferencing. We acknowledge the dual-use nature of this work and the ethical sensitivities inherent in any study involving human subjects and potential vulnerabilities. From its inception, this project was designed with a proactive ethical framework to ensure our methods were sound, our participants were protected, and our findings would contribute positively to the security community. All procedures were conducted in strict accordance with the ethical principles of the Belmont Report and our institution's policies.

\subsection{Study Justification and Societal Benefit} The primary motivation for this research is to proactively identify and understand a previously unknown privacy risk in ubiquitous video conferencing systems. By demonstrating the feasibility of the ``FaceTell'' attack, our goal is to alert developers, platform providers, and the public to this threat. The principal societal benefit of this work is to enable the timely development of countermeasures, thereby protecting millions of users from potential eavesdropping attacks. We argue that the benefit of raising awareness and fostering defensive innovation significantly outweighs the risks associated with the disclosure of this attack method.

\subsection{Institutional Review Board (IRB) Approval}
We have prioritized ethical considerations at every stage of our research. 
The complete research protocol---encompassing participant recruitment, the informed consent process, data management, and our privacy protection plan---was formally reviewed and approved by our institution's Institutional Review Board (IRB).

\subsection{Data Minimization and Protection} 
Our data handling practices were guided by the principles of data minimization and robust security. 
\begin{itemize} 
\item \textbf{Minimization}: We collected only the data essential for validating our attack model---specifically, facial video feeds during scripted application use. No audio was recorded, and no personal or personally identifiable information beyond the facial data itself was collected. 
\item \textbf{Anonymity and Protection}: To safeguard participant privacy, all collected video data was immediately de-identified by assigning a non-identifiable participant ID. The raw and processed data were stored on encrypted, access-controlled servers, with access strictly limited to the core research team. 
Furthermore, to rigorously protect the privacy of our volunteer participants, all facial images illustrated in this publication are from the authors.
\end{itemize} 

\subsection{Risk/Benefit Analysis and Responsible Disclosure}

We detail the risks and benefits associated with this study by answering the following questions:

\textbf{1) Who may benefit from the FaceTell attack and how they will be harmed by the disclosure of the proposed countermeasures?}

Malicious actors, such as hackers, would have benefited from the undefended state of the FaceTell attack. 
Hackers could use this attack to gain unauthorized access to sensitive information, such as financial data, personal identities, and business secrets. 

With the disclosure of the countermeasures proposed in~\cref{subsec-countermeasures}, these malicious actors will find it much more difficult to execute such attacks. 
They will have to invest significant time and resources to find new ways to bypass these defenses. 

\textbf{2) Who may be harmed by the FaceTell attack and what are their benefits from the countermeasures?}

Millions of individual users who participate in video conferences are at risk of having their private information eavesdropped on. 
This includes personal interests, health information, and political inclination. 

The development of countermeasures (described in~\cref{subsec-countermeasures}) will protect their privacy and prevent potential identity theft and financial fraud. 
They can participate in video conferences with greater peace of mind, knowing that their multitasking behaviors are unknown to others.

\textbf{3) What may need to change for the countermeasures to be implemented and how will that affect people?}

Three countermeasures are proposed in~\cref{subsec-countermeasures}.
We analyze each countermeasure separately.
\begin{itemize}
    \item \textbf{Introducing Additional Dynamic Light Sources.} 
    Users need to place an extra dynamic light source, e.g., a second screen, outside the camera's FoV. 
    This setup should be easy and low-cost. However, more screens may accelerate visual fatigue during online meetings.
    \item \textbf{Enabling Skin Smoothing Filters.} 
    Users just have to turn on filtering features during video conferencing. While these features are easy to use, some video conferencing participants may be reluctant to use them as the smoothed faces may cause issues relating to authenticity.
    \item \textbf{Adding Adversarial Perturbations.} 
    The owners of video conferencing platforms need to conduct significant technical development to develop and integrate this feature.
    Once developed, it can be an automatic defense. 
    But the development takes time and may consume extra computational resources, impacting device performance and power consumption.
\end{itemize}

In conclusion, the publication of our research on the FaceTell attack will encourage the development of countermeasures that protect millions of users from potential eavesdropping attacks. 
While there are risks associated with the disclosure of the attack method, the long-term benefits of raising awareness and fostering defensive innovation clearly outweigh these risks.

\subsection{Informed Consent Documentation}
We recruited 24 adult volunteers who provided their written informed consent. To ensure transparency and full understanding of the risks, participants were required to review and affirm a detailed pre-experimental questionnaire. The specific items affirmed by participants were as follows:
\begin{enumerate}[nosep, itemsep=1.9pt, leftmargin=*]
    \item Do you confirm that your participation is entirely voluntary and that you are an adult providing your own consent?
    \item Do you consent to the recording of your facial video stream while using specific applications on the computer during video conferencing?
    \item Do you confirm that no private audio, keyboard input, or personally identifiable information (PII) other than your visual presence was recorded?
    \item Do you understand that all raw video data will be immediately de-identified (assigned a non-identifiable ID) and stored on an encrypted, access-controlled server?
    \item Do you understand that the full, sensitive video dataset collected from the subjects will not be made public?
    \item Do you authorize the use of cropped, processed, and low-resolution images of your face in the publication figures for scientific illustration?
    \item Do you understand that the primary purpose of using these images is to demonstrate the visual patterns of light reflection and not to reveal your personal identity?
    \item Do you understand that consenting to the use of your images will contribute to the validation of a crucial defensive security finding (in accordance with the Beneficence principle)?
    \item Do you understand that we cannot use your images for other purposes without your consent?
    \item Do you understand that it is our responsibility to maintain your data integrity and protect it from security threats?
    \item Do you understand that we will inform you in a timely manner if any security issues affect your data?
    \item Do you understand that you have the right to withdraw from the study at any time, for any reason, without suffering penalty or loss of benefits?
\end{enumerate}

\cleardoublepage

\section*{Open Science}
\setcounter{subsection}{0}
We are committed to the principles of open and reproducible research. 
To support the verification of FaceTell and allow other researchers to build upon our work, we provide a comprehensive artifact package that includes our source code, pre-trained models, and a sample dataset. 

\subsection{Artifact Availability} 
Our artifacts are available for review via a Zenodo link. 
This package allows for the replication of FaceTell as described in the paper. 
The link is: \url{https://zenodo.org/records/16925324}. 

\subsection{Artifact Description} 
The provided artifact package is organized into the following components, with detailed instructions available in the \texttt{README.md} file: 
\begin{itemize} 
\item \textbf{Source Code}: We provide all Python scripts for the two main components of our pipeline: \begin{enumerate} 
\item \textit{Face Segmentation and Reconstruction}: It includes scripts for video processing, face detection, and super-resolution. 
\item \textit{Screen Content Inference}: It includes the complete code for training our two-tier classification model, for testing new data using the provided pre-trained models, and for the heuristic label correction algorithm. 
\end{enumerate} 
\item \textbf{Pre-trained Models}: To facilitate immediate testing and verification of FaceTell, we provide the full set of pre-trained models described in the paper. 
This includes the Tier-1 category discriminator and all specialized Tier-2 application predictors.
\item \textbf{Dataset}: The full dataset, collected from 24 human subjects, contains sensitive facial video information. 
In accordance with our IRB-approved protocol and our ethical commitment to protect participant privacy, the full dataset cannot be made publicly available. 
To address this limitation while still supporting reproducibility, we provide a sample dataset. 
This dataset contains data from a single participant who fully understands the study's implications and has explicitly consented to their data being used for this specific purpose. 
This sample data is structured identically to our full dataset and is sufficient to run the entire code pipeline, from data processing to final classification, allowing reviewers to verify the functionality and methodology of FaceTell. 
\end{itemize} 

\subsection{Reproducibility Statement} 
We believe that the combination of our detailed paper, the complete source code, the pre-trained models, and the sample dataset provides sufficient materials for the research community to verify our core claims and reproduce our methodology.
\cleardoublepage
\bibliographystyle{plain}
\bibliography{sample}

@String{Computer = "{IEEE} Computer" }

@String{Springer = "Springer-Verlag" }

@misc{Kumar2024,
  author ={Naveen Kumar},
  title ={Zoom User Statistics 2024 — Market Share \& Revenue},
  year ={2024},
  note =         {\url{http://pages.cs.wisc.edu/~remzi/OSTEP/}}
}

@article{riedl2022stress,
  author={Riedl, Ren{\'e}},
  title={On the stress potential of videoconferencing: definition and root causes of {Zoom} fatigue},
  journal={Electronic Markets},
  volume={32},
  number={1},
  pages={153--177},
  year={2022},
  publisher={Springer},
  note ={\url{https://link.springer.com/article/10.1007/s12525-021-00501-3}}  
}

@inproceedings{backes2008compromising,
  author={Backes, Michael and D{\"u}rmuth, Markus and Unruh, Dominique},
  title={Compromising reflections-or-how to read {LCD} monitors around the corner},
  booktitle={Proceedings of the IEEE Symposium on Security and Privacy (SP)},
  pages={158--169},
  year={2008},
  note = {\url{https://ieeexplore.ieee.org/document/4531151}}
}

@inproceedings{backes2009tempest,
  author={Backes, Michael and Chen, Tongbo and D{\"u}rmuth, Markus and Lensch, Hendrik PA and Welk, Martin},
  title={Tempest in a teapot: Compromising reflections revisited},
  booktitle={Proceedings of the IEEE Symposium on Security and Privacy (SP)},
  pages={315--327},
  year={2009},
  note = {\url{https://ieeexplore.ieee.org/document/5207653}}
}

@inproceedings{raguram2011ispy,
  author={Raguram, Rahul and White, Andrew M and Goswami, Dibyendusekhar and Monrose, Fabian and Frahm, Jan-Michael},
  title={{iSpy}: automatic reconstruction of typed input from compromising reflections},
  booktitle={Proceedings of the ACM conference on Computer and Communications Security (CCS)},
  pages={527--536},
  year={2011},
  note = {\url{https://dl.acm.org/doi/10.1145/2046707.2046769}}
}

@inproceedings{xu2013seeing,
  author={Xu, Yi and Heinly, Jared and White, Andrew M and Monrose, Fabian and Frahm, Jan-Michael},
  title={Seeing double: Reconstructing obscured typed input from repeated compromising reflections},
  booktitle={Proceedings of the ACM conference on Computer and Communications Security (CCS)},
  pages={1063--1074},
  year={2013},
  note = {\url{https://dl.acm.org/doi/10.1145/2508859.2516709}}
}

@inproceedings{long2023private,
  author={Long, Yan and Yan, Chen and Xiao, Shilin and Prasad, Shivan and Xu, Wenyuan and Fu, Kevin},
  title={{Private Eye}: On the limits of textual screen peeking via eyeglass reflections in video conferencing},
  booktitle={Proceedings of the IEEE Symposium on Security and Privacy (SP)},
  pages={3432--3449},
  year={2023},
  note = {\url{https://www.computer.org/csdl/proceedings-article/sp/2023/933600a870/1OXGUMtuJLa}}
}

@article{egger20203d,
  author={Egger, Bernhard and Smith, William AP and Tewari, Ayush and Wuhrer, Stefanie and Zollhoefer, Michael and Beeler, Thabo and Bernard, Florian and Bolkart, Timo and Kortylewski, Adam and Romdhani, Sami and others},
  title={{3D} morphable face models—past, present, and future},
  journal={ACM Transactions on Graphics},
  volume={39},
  number={5},
  pages={1--38},
  year={2020},
  note =         {\url{https://dl.acm.org/doi/10.1145/3395208}}
}

@article{svilainis2008led,
  author={Svilainis, Linas},
  title={{LED} directivity measurement in situ},
  journal={Measurement},
  volume={41},
  number={6},
  pages={647--654},
  year={2008},
  publisher={Elsevier},
  note =         {\url{https://www.sciencedirect.com/science/article/abs/pii/S0263224107000917}}
}

@Techreport{radiometry2001,
  author =       "",
  year =         "2001",
  title =        "The Radiometry of Light Emitting Diodes",
  institution =  "Labsphere Inc.",
  type =         "Technical Guide",
  number =       "",
  address =      "",
  month =        "",
  note =         {\url{https://www.labsphere.com/wp-content/uploads/2021/09/Radiometry-of-Light-Emitting-Diodes.pdf}}
}

@incollection{phong1998illumination,
  title={Illumination for computer generated pictures},
  author={Phong, Bui Tuong},
  booktitle={Seminal graphics: pioneering efforts that shaped the field},
  pages={95--101},
  year={1998},        
  note = {\url{http://www.cs.northwestern.edu/~ago820/cs395/Papers/Phong_1975.pdf}}
}

@article{svilainis2010numerical,
  title={Numerical comparison of {LED} directivity approximation functions for video displays},
  author={Svilainis, Linas and Dumbrava, Vytautas},
  journal={Displays},
  volume={31},
  number={4-5},
  pages={196--204},
  year={2010},
  publisher={Elsevier},
  note =         {\url{https://www.sciencedirect.com/science/article/abs/pii/S0141938210000661}}
  
}

@article{al2011illumination,
  title={Illumination normalization of facial images by reversing the process of image formation},
  author={Al-Osaimi, Faisal R and Bennamoun, Mohammed and Mian, Ajmal},
  journal={Machine Vision and Applications},
  volume={22},
  number={6},
  pages={899--911},
  year={2011},
  publisher={Springer},
  note =         {\url{https://link.springer.com/article/10.1007/s00138-010-0309-5}}
}

@article{cecconello2019skype,
  title={{Skype \& Type}: Keyboard eavesdropping in {Voice-over-IP}},
  author={Cecconello, Stefano and Compagno, Alberto and Conti, Mauro and Lain, Daniele and Tsudik, Gene},
  journal={ACM Transactions on Privacy and Security},
  volume={22},
  number={4},
  pages={1--34},
  year={2019},
  note =         {\url{https://dl.acm.org/doi/10.1145/3365366}}
}

@inproceedings{he2017mask,
  author={He, Kaiming and Gkioxari, Georgia and Doll{\'a}r, Piotr and Girshick, Ross},
  title={Mask {R-CNN}},
  booktitle={Proceedings of the IEEE International Conference on Computer Vision (ICCV)},
  pages={2961--2969},
  year={2017},
  note =         {\url{https://openaccess.thecvf.com/content_ICCV_2017/papers/He_Mask_R-CNN_ICCV_2017_paper.pdf}}
}

@inproceedings{viola2001rapid,
  author={Viola, Paul and Jones, Michael},
  title={Rapid object detection using a boosted cascade of simple features},
  booktitle={Proceedings of the IEEE Conference on Computer Vision and Pattern Recognition (CVPR)},
  year={2001},
  pages={I--I},
  note =         {\url{https://ieeexplore.ieee.org/document/990517}}
}

@inproceedings{wang2024camixersr,
  author={Wang, Yan and Liu, Yi and Zhao, Shijie and Li, Junlin and Zhang, Li},
  title={{CAMixerSR}: Only Details Need More "Attention"}, 
  booktitle={Proceedings of the IEEE Conference on Computer Vision and Pattern Recognition (CVPR)},
  pages={25837--25846},
  year={2024},
  note =         {\url{https://openaccess.thecvf.com/content/CVPR2024/papers/Wang_CAMixerSR_Only_Details_Need_More_Attention_CVPR_2024_paper.pdf}}
}

@inproceedings{he2016deep,
  author={He, Kaiming and Zhang, Xiangyu and Ren, Shaoqing and Sun, Jian},
  title={Deep residual learning for image recognition},
  booktitle={Proceedings of the IEEE Conference on Computer Vision and Pattern Recognition (CVPR)},
  pages={770--778},
  year={2016},
  note =         {\url{https://ieeexplore.ieee.org/document/7780459}}
}

@inproceedings{woo2018cbam,
  author={Woo, Sanghyun and Park, Jongchan and Lee, Joon-Young and Kweon, In So},
  title={{CBAM}: Convolutional block attention module},
  booktitle={Proceedings of the European Conference on Computer Vision (ECCV)},
  pages={3--19},
  year={2018},
  note =         {\url{https://openaccess.thecvf.com/content_ECCV_2018/papers/Sanghyun_Woo_Convolutional_Block_Attention_ECCV_2018_paper.pdf}}
}

@inproceedings{Daniel2019acoustic,
  author={Genkin, Daniel and Pattani, Mihir and Schuster,  Roei and Tromer, Eran},
  title={Synesthesia: Detecting Screen Content via Remote Acoustic Side Channels},
  booktitle={Proceedings of the IEEE Symposium on Security and Privacy (SP)},
  pages={853--869},
  year={2019},
  note =         {\url{https://www.computer.org/csdl/proceedings-article/sp/2019/666000a853/1dlwjC6dIT6}}
}

@inproceedings{hayashi2014EM,
  author={Hayashi, Yuichi and Homma, Naofumi and Miura, Mamoru and Aoki, Takafumi and Sone, Hideaki},
  title={A Threat for Tablet {PCs} in Public Space: Remote Visualization of Screen Images Using EM Emanation},
  booktitle={Proceedings of the ACM Conference on Computer and Communications Security (CCS)},
  pages={954--965},
  year={2014},
  note =         {\url{https://dl.acm.org/doi/10.1145/2660267.2660292}}
}

@inproceedings{ni2023fingerprints,
  author={Ni, Tao and Zhang, Xiaokuan and Zhao, Qingchuan},
  title={Recovering Fingerprints from In-Display Fingerprint Sensors via Electromagnetic Side Channel},
  booktitle={Proceedings of the ACM Conference on Computer and Communications Security (CCS)},
  pages={253--267},
  year={2023},
  note =         {\url{https://dl.acm.org/doi/10.1145/3576915.3623153}}
}

@inproceedings{li2020wavespy,
  author={Li, Zhengxiong and Ma, Fenglong and Rathore, Aditya Singh and Yang, Zhuolin and Chen, Baicheng and Su, Lu and Xu, Wenyao},
  title={WaveSpy: Remote and Through-wall Screen Attack via {mmWave} Sensing},
  booktitle={Proceedings of the IEEE Symposium on Security and Privacy (SP)},
  pages={217-232},
  year={2020},
  note =         {\url{https://ieeexplore.ieee.org/document/9152804}}
}

@inproceedings{ni2023applistener,
  author={Ni, Tao and Lan, Guohao and Wang, Jia and Zhao, Qingchuan and Xu, Weitao},
  title={Eavesdropping mobile app activity via radio-frequency energy harvesting},
  booktitle={Proceedings of the USENIX Security Symposium (USENIX Security)},
  pages={3511-3528},
  year={2023},
  note =         {\url{https://www.usenix.org/system/files/usenixsecurity23-ni.pdf}}
}

@inproceedings{fang2018csi,
  author={Fang, Song and Markwood, Ian and Liu, Yao and Zhao, Shangqing and Lu, Zhuo and Zhu, Haojin},
  title={No Training Hurdles: Fast Training-Agnostic Attacks to Infer Your Typing},
  booktitle={Proceedings of the ACM Conference on Computer and Communications Security (CCS)},
  pages={1747–1760},
  year={2018},
  note =         {\url{https://dl.acm.org/doi/10.1145/3243734.3243755}}
}

@article{balzer2010principles,
  title={Principles of shape from specular reflection},
  author={Balzer, Jonathan and Werling, Stefan},
  journal={Measurement},
  volume={43},
  number={10},
  pages={1305--1317},
  year={2010},
  publisher={Elsevier},
  note =         {\url{https://www.sciencedirect.com/science/article/abs/pii/S0263224110001570}}
}

@article{goodfellow2014explaining,
  title={Explaining and harnessing adversarial examples},
  author={Goodfellow, Ian J and Shlens, Jonathon and Szegedy, Christian},
  journal={arXiv:1412.6572},
  year={2014},
  note =         {\url{https://arxiv.org/pdf/1412.6572}}
}

@inproceedings{wasswa2022proof,
  title={The proof is in the glare: On the privacy risk posed by eyeglasses in video calls},
  author={Wasswa, Hassan and Serwadda, Abdul},
  booktitle={Proceedings of the 2022 ACM on International Workshop on Security and Privacy Analytics},
  pages={46--54},
  year={2022},
  note ={\url{https://dl.acm.org/doi/abs/10.1145/3510548.3519378}}
}

@inproceedings{weinberg2011still,
  title={I still know what you visited last summer: Leaking browsing history via user interaction and side channel attacks},
  author={Weinberg, Zachary and Chen, Eric Y and Jayaraman, Pavithra Ramesh and Jackson, Collin},
  booktitle={2011 IEEE Symposium on Security and Privacy},
  pages={147--161},
  year={2011},
  note ={\url{https://ieeexplore.ieee.org/document/5958027}}
}

@misc{firstpagesage2022session,
  author = {Evan Bailyn},
  title = {{Average Session Duration By Industry}},
  year = {2022},
  note = {\url{https://firstpagesage.com/reports/average-session-duration-by-industry/}}
}

@misc{daniel2025digital,
  author = {Daniel Porter},
  title = {{Digital Devices and Your Eyes}},
  year = {2025},
  note = {\url{https://www.aao.org/eye-health/tips-prevention/digital-devices-your-eyes}}
}

@misc{toolbox2004illuminace,
  author = {{The Engineering ToolBox}},
  title = {{Illuminance - Recommended Light Levels}},
  year = {2004},
  note = {\url{https://www.engineeringtoolbox.com/light-level-rooms-d_708.html}}
}

@article{dias2017toward,
  title={Toward Proper Evaluation of Light Dose in Indoor Office Environment by Frontal Lux Meter},
  author={Dias, Ma{\'i}ra Vieira and Motamed, Ali and Scarazzato, Paulo Sergio and Scartezzini, Jean-Louis},
  journal={Energy Procedia},
  year={2017},
  pages={835--840},
  volume={122},
  publisher={Elsevier},
  note = {\url{https://www.sciencedirect.com/science/article/pii/S187661021733148X?ref=pdf_download&fr=RR-2&rr=9ac48880bc91a9b5}}
}

@article{zhang2022study,
  title={Study of Human Visual Comfort Based on Sudden Vertical Illuminance Changes},
  author={Zhang, Jiuhong and Lv, Kunjie and Zhang, Xiaoqian and Ma, Mingxiao and Zhang, Jiahui},
  journal={Buildings},
  volume={12},
  number={8},
  pages={1127},
  year={2022},
  publisher={MDPI},
  note = {\url{https://www.mdpi.com/2075-5309/12/8/1127}}
}

\cleardoublepage
\appendix

\section{Details of Implementation}

\subsection{Screenshots of Selected Applications}

 \begin{figure}[h]
    \centering
    \includegraphics[width=0.9\linewidth]{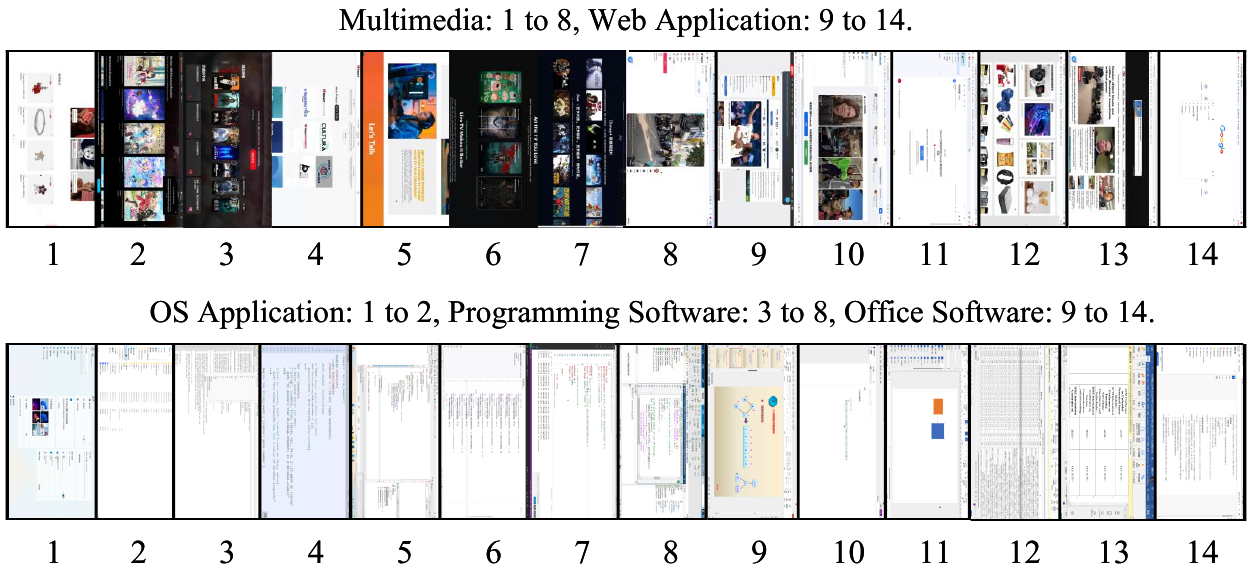}
    \caption{Screenshots of 28 selected applications. Each image is
 rotated by 90 degrees and concatenated horizontally.}
    \label{fig-application_category}
\end{figure} 

\subsection{Subject Information}

\begin{table}[!h]
\centering
\caption{Subject information.}
\label{tab:user_information}
\footnotesize  
\setlength{\tabcolsep}{1.7pt} 
\begin{tabular}{*{6}{c}} 
\toprule
\textbf{Subject} & \textbf{Gender} & \textbf{Facial Occlusion} & \textbf{Platform} & \textbf{Computer} & \textbf{Location} \\ 
\midrule
1  & Female & None       & WeChat          & ASUS   & Bedroom 1  \\
2  & Female & None       & WeChat          & Lenovo & Bedroom 2  \\
3  & Female & None       & Skype           & HP     & Office 1   \\
4  & Female & Eyeglasses & Zoom            & ASUS   & Office 2   \\
5  & Female & Eyeglasses & Zoom            & ASUS   & Office 3   \\
6  & Female & Eyeglasses & Skype           & Lenovo & Bedroom 3  \\
7  & Female & Eyeglasses & Skype           & Lenovo & Office 4   \\
8  & Female & Eyeglasses & Teams & HP     & Office 5   \\
9  & Female & Mask       & Teams & HP     & Dormitory 1\\
10 & Female & Mask       & WeChat          & ASUS   & Office 2   \\
11 & Male   & None       & Teams & Lenovo & Office 2   \\
12 & Male   & None       & Zoom            & ASUS   & Office 1   \\
13 & Male   & Eyeglasses & WeChat          & HP     & Office 1   \\
14 & Male   & Eyeglasses & WeChat          & Lenovo & Office 1   \\
15 & Male   & Eyeglasses & WeChat          & HP     & Office 6   \\
16 & Male   & Eyeglasses & Skype           & Lenovo & Dormitory 2\\
17 & Male   & Eyeglasses & Skype           & HP     & Office 6   \\
18 & Male   & Eyeglasses & Zoom            & Lenovo & Office 6   \\
19 & Male   & Eyeglasses & Zoom            & ASUS   & Office 1   \\
20 & Male   & Eyeglasses & Teams & ASUS   & Dormitory 3\\
21 & Male   & Eyeglasses & Teams & HP     & Office 2   \\
22 & Male   & Eyeglasses & Teams & ASUS   & Office 7   \\
23 & Male   & Mask       & Skype           & Lenovo & Office 2   \\
24 & Male   & Mask       & Zoom            & HP     & Office 1   \\
\bottomrule
\end{tabular}
\end{table}

\section{Details of Evaluations}

\subsection{Parameter Determination}

\begin{figure}[h]
    \centering
    \includegraphics[width=0.9\linewidth]{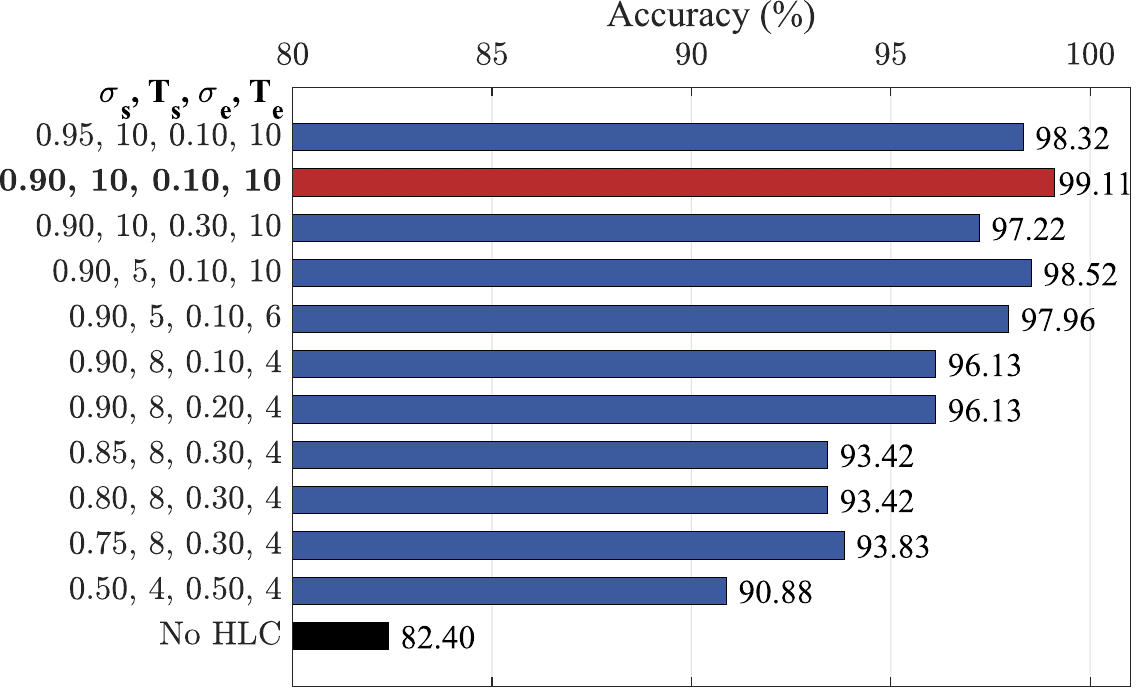}
    \caption{The multimedia predictor's performance under different HLC parameter settings.}
    \label{fig-accuracy_comparison}
\end{figure}

We justify the choice of the four parameters in the proposed heuristic label correction (HLC) algorithm.
There are four key hyperparameters, i.e., $\sigma_s$, $T_s$, $\sigma_e$, and $T_e$.
$T_s$ and $\sigma_s$ are the window length and decision threshold of $\text{Start}(\cdot)$, and $T_e$ and $\sigma_e$ are those of $\text{End}(\cdot)$.
We evaluate our multimedia predictor, which has the most prediction labels, on the testing dataset.
As depicted in Figure~\ref{fig-accuracy_comparison}, the HLC algorithm can significantly improve the predictor's performance in each setting.
Moreover, the larger $\sigma_s$ and $T_s$ are, the better the performance is generally achieved.
While a small $\sigma_e$ and a large $T_e$ can also boost classification performance in most cases.
After dozens of rounds of tests, we find that the highest accuracy of 99.11\% is obtained in our experiments, when \(\sigma_s=0.90\), \(T_s=10\), \(\sigma_e=0.10\), and \(T_e=10\).

\subsection{Impact of Video Conferencing Platform and Computer Hardware} 
\begin{figure}[h]
    \centering
    \includegraphics[width=0.8\linewidth]{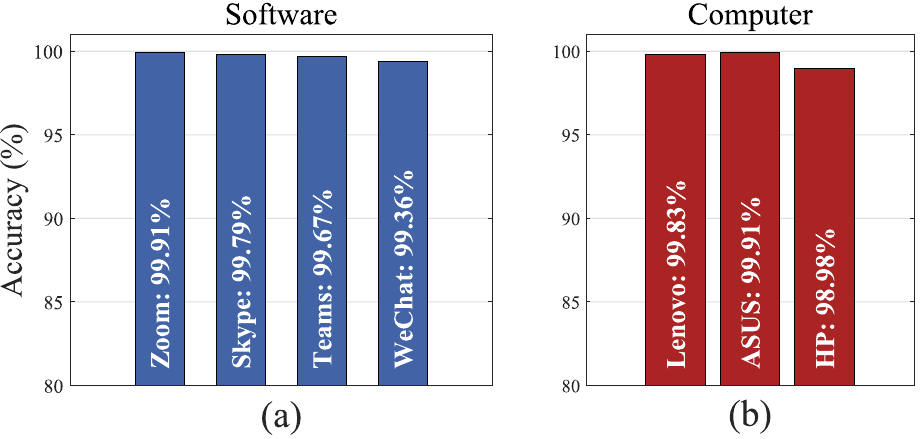}
    \caption{Performance under different video conferencing platforms and computer hardware.}
    \label{fig-software_and_computer_brand}
\end{figure}

We investigate the impact of two key external factors: variations in video conferencing platforms and computer hardware. 
In our experiments, four mainstream video conferencing platforms, i.e., Skype, Zoom, Teams, and WeChat, run on three laptops with different brands, including ASUS, Lenovo, and HP.
As shown in Figure~\ref{fig-software_and_computer_brand}~(a), FaceTell does not show significant performance differences on Zoom, Skype, Teams, and WeChat, and the accuracy on each platform reaches over 99\%.
The same observation can also be found in Figure~\ref{fig-software_and_computer_brand}~(b), where the prediction accuracy on different computer hardware is around 99\%. 
The above results indicate that FaceTell has good stability and generalization capability on different software and hardware platforms.

\end{document}